\documentclass{article}

\usepackage[utf8]{inputenc} 
\usepackage{amsmath}
\usepackage{amsthm}
\usepackage{amsfonts}
\usepackage{amssymb}
\usepackage{epsfig,color,mathrsfs}
\usepackage{enumerate}
\usepackage{hyperref,cancel,soul}
\hypersetup{colorlinks=true,hidelinks=true}
\usepackage[titletoc]{appendix}
\usepackage{multirow}
\usepackage{wrapfig}
\usepackage{graphicx}
\usepackage[usenames,dvipsnames]{xcolor}

\usepackage{tikz}
\usetikzlibrary{arrows,automata,positioning}
\usepackage{wrapfig}
\usepackage{subcaption}
\usepackage{symbols}

\usepackage{framed}
\usepackage[boxed,ruled,vlined,linesnumbered]{algorithm2e}
\usepackage{marginnote}
\usepackage{geometry}

\newtheorem{theorem}{Theorem}
\newtheorem{lemma}[theorem]{Lemma}
\newtheorem{corollary}[theorem]{Corollary}
\newcommand{\Section}{\section}

\newcommand{\Subsection}{\subsection}
\newcommand{\Subsubsection}[1]{\noindent {\bf #1.}~~}

\newcommand{\ch}{\mathcal{H}}
\newcommand{\cV}{{\mathtt{V}}}
\newcommand{\cE}{\mathtt{E}}
\newcommand{\rp}{\mathcal{RP}}

\newcommand{\dgraph}{\mathscr{G}}

\newcommand{\cs}{\mathcal{CS}}

\newcommand{\toFix}[1]{}
\newcommand{\remove}[1]{}
\newcommand{\short}[1]{}
\newcommand{\alternative}[1]{}
\newcommand{\hide}[1]{}

\begin{document}

\title{Shared-object System Equilibria:\\ Delay and Throughput Analysis\\ \large{(Technical Report)}}


%
\author{
%
%
Iosif~Salem~~~~~ Elad~M.~Schiller~~~~~ Marina~Papatriantafilou~~~~~ 
Philippas~Tsigas\\~\\
Department of Computer Science and Engineering\\
         Chalmers University of Technology\\
         S--412 96, G\"oteborg, Sweden\\
      \texttt{\{iosif, elad, ptrianta, tsigas\}@chalmers.se}
}
\date{}

\maketitle
\begin{abstract}
We consider shared-object systems that require their threads to fulfill the system jobs by first acquiring sequentially the objects needed for the jobs and then holding on to them until the job completion. Such systems are in the core of a variety of shared-resource allocation and synchronization systems.
This work opens a new perspective to study the expected job delay and throughput analytically, given the possible set of jobs that may join the system dynamically.
We identify the system dependencies that cause contention among the threads as they try to acquire the job objects. We use these observations to define the shared-object system equilibria. We note that the system is in equilibrium whenever the rate in which jobs arrive at the system matches the job completion rate. These equilibria consider not only the job delay but also the job throughput, as well as the time in which each thread blocks other threads in order to complete its job. We then further study in detail the thread work cycles and, by using a graph representation of the problem, we are able to propose procedures for finding and estimating equilibria, i.e., discovering the job delay and throughput, as well as the blocking time. To the best of our knowledge, this is a new perspective, that can provide better analytical tools for the problem, in order to estimate performance measures similar to ones that can be acquired through experimentation on working systems and simulations, e.g., as job delay and throughput in (distributed) shared-object systems.
\end{abstract}

\Section{Introduction}
We consider shared-object systems that require their threads to fulfill the system jobs by first acquiring sequentially all of the job objects. The job then holds on to these objects until the job operation is done. We identify the system dependencies that cause contention among the threads as they try to acquire the job objects.
We study the (stochastic) processes of job arrival and completion with an emphasis on the cases in which the job arrival rate matches the job completion rate, i.e., the job throughput. In these cases, the system is in a {\em shared-Object System Equilibrium} (OSE). For a given $\varepsilon > 0$ and an OSE, we say that the system is in an {\em $\varepsilon$-OSE} when the completion rate of any job differs from the one of an OSE by at most $\varepsilon$.
We study the conditions for a given shared-object system to be in an OSE as well as contention-related properties of OSEs, i.e., the {\em expected} job delay and completion rate, as well as the time in which each thread blocks other threads and by that prevents them from making progress.
We propose an analytical procedure for finding (in polynomial time) $\varepsilon$-OSEs. Moreover, we estimate the performance measures of 
 systems that are in $\varepsilon$-OSE.
%
%

The existing practice considers job delay and completion rate as the performance measures of working systems. Empirical experiments often study shared-resource systems at their saturation point in which the system is at its peak utilization. 
Let us describe peak utilization scenarios using two vectors; one for job arrival rates and another for their completion rates. A saturation point is the case in which: (1) the system is in equilibrium, i.e., the arrival rate of any particular job matches the completion rate of this job, as well as (2) the system is at the stage at which a higher arrival rate of any job to the system cannot increase its completion rate.     
Our study considers the entire range of these equilibria rather than just peak utilization scenarios (Section~\ref{s:preliminaries}). 
We then propose a procedure for finding $\varepsilon$-OSEs, if such exist in the given system (Section~\ref{s:probDesc}). Once we find an $\varepsilon$-OSE, we can estimate its performance measures, i.e., job delay, completion rate and blocking time.
To this end, we develop a number of analytical tools for OSEs.
%
%
%
%
Given the job arrival rates, we show how to estimate the probabilities for threads to follow a certain object acquisition sequence (Section~\ref{s:RoutingBlocking}). We are then able to formulate recursive equations (with interdependencies) for calculating the blocking periods and the completion rates (sections~\ref{s:blocking}, and respectively,~\ref{s:throughput}). We overcome these dependencies and solve these recursive equations by analysing the thread work cycles (Section~\ref{s:cont_graphs}).
%
%

\Subsubsection{Related Work}
Our problem domain considers computing entities, which are called threads. Each thread runs a sequential program that has to acquire reusable resources (objects), often several at the same time, for a bounded time of use. To guarantee deadlock absence, it is important that all threads acquire the objects in an ordered manner. For example, one can deterministically define a partial order among the objects, such that the threads acquire them in totally ordered manner. 
%
%
We consider a generalization of the dining philosophers problem, as in~\cite{DBLP:books/mk/Lynch96, DBLP:conf/opodis/PapatriantafilouT97}, in which every job includes a fixed set  of objects that it may need.
This problem has well-known results studying the worst-case job delays, which  may even be exponential on metrics, such as the chromatic number of the resource graph~\cite{DBLP:journals/jcss/Lynch81, DBLP:books/mk/Lynch96}. In this graph, the vertices (objects) are connected if there is at least one thread that may request them both at any point in time.
In the context of actual systems, the expected time is rather different than the worst case and therefore computer experiments are the common way for evaluating the system performance.
%
%
We provide a new perspective that enables an analysis of the evaluation metrics
%
%
by considering measures both at the system level and at the level of each resource. 
In particular, we consider performance measures that are associated with each resource, such as the delay, completion rate and blocking time. On the system level, we consider the job arrival and completion rates, as well as the total number of threads, $N$, and objects, $M$.

\Subsubsection{Our contribution}
We study analytical tools that provide the means to estimate performance measures of working distributed systems. In the context of synchronization challenges that are modeled via a generalization of the dynamic dining philosophers problem, our analytical tools are the first, to the best of our knowledge, to consider performance measures similar to the ones that can be acquired via experimentation on working systems and simulations. 

For a given number of threads and job arrival rates, we provide a way to analyze the delay of jobs and their completion rates as well as the time for which the threads are blocked.
In addition to the job completion period (Lemma \ref{lemma:job_completion}), we analyze a number of key properties, such as the probability to request a particular resource after the acquisition of another specific resource, 
the time during which threads that have acquired such a particular resource block other threads that ask to access the same resource (Lemma~\ref{lemma:blocking}) as well as the time between two requests to access such resources (Lemma~\ref{lemma:agg_throughput}). 
Since these properties have interdependencies due to thread blocking, we show how the concept of thread work cycles can be represented in subsystems that also include such interdependencies but have no thread blocking. 
This way, we can resolve these interdependencies (Theorem~\ref{thm:BaynatDallery}) and estimate the performance of the given (distributed) shared resource system. 
Moreover, we use the work cycle events (sections~\ref{ss:work_cycles} and \ref{ss:cond-consec}) to verify our modeling approach.
We present a procedure for satisfying approximately the equilibrium conditions and by that find an $\varepsilon$-OSE as well as the performance measures of the studied system (Section~\ref{s:probDesc}).

Our contribution can facilitate early-stage evaluations of systems that are similar to the studied one.
Moreover, using our proposed methods, one can analytically, rather than via empirical experiments, study trade-offs among OSEs. Such trade-offs can facilitate the design of mechanisms for adjusting the number of threads and job arrival rates according to the performance measures of a dynamic system.


\Section{Preliminaries}
\label{s:preliminaries}
We consider a system that includes {\em (system) items}, which are {\em (totally ordered) objects}, $(object[1]$, $\ldots$, $object[M])$, and {\em (totally ordered) threads}, $(thread[1]$, $\ldots$, $thread[N])$. The objects are shared in a mutually exclusive way, i.e., only one thread at a time may gain access to an object. Each thread is to carry out one {\em job} at a time, where $job_i$ $= \langle objs_i$, $operation_i \rangle$, $J$ is the number of the system's jobs, $i \in [1, J]$ and $objs_i = ( object_{i_1}$, $\ldots$, $object_{i_k} )$ is 
an arbitrary, 
non-empty subsequence of $(object[1], \ldots, object[M])$, and thus $objs_i$ follows the same order. 
Note that we assume that $objs_i$ is a fixed vector and that different jobs may have different object vectors of different lengths. 
%
Moreover, the {\em (job) operation time}, $O_{i}$, is a random variable with a known distribution.
%
Namely, we assume that the time it takes to execute the job operation is provided, say, via a profiler.

\Subsection{Acquisition paths, periods and requests}
Suppose that the system assigns $job_i$ to $thread[n]:1\leq n \leq N$. In this case, $job_i$'s {\em (acquisition) path} is the vector, $path_{i,n}$ $=$ $( thread[n]$, $object[i_1]$, $\ldots$, $object[i_k] )$, in which $thread[n]$ carries out $job_i$'s operation after it has sequentially acquired $object_{i_1}$, $\ldots$, $object_{i_k}$.
A thread can acquire a particular object, $object[i]$, by pending its {\em (acquisition) request} in a (first in, first out) queue $Q(object[i])$ until all (previously) waiting threads in $Q(object[i])$ have acquired and released $object[i]$.
The {\em acquisition period}, $A$, is a known random variable that refers to a period that starts when a thread has acquired an object (or just been assigned to a new job) and ends as soon as that thread places a request for the next object. 
Namely, we assume that the time it takes to send a request after a supply event is provided, say, via a profiler. 
Once the thread sequentially acquires the entire object set, $object_{i_1},$ $\ldots,$ $object_{i_k}$, it executes the job operation, $operation_i$, before completing the job.
We say that a thread is {\em blocking} when other threads are queuing for its acquired objects. That happens whenever different jobs have overlapping object vectors. Note however that threads carry out jobs within finite time even in the presence of blocking, because our definition of acquisition paths considers object acquisition according to a common (total) order.
This work focuses on systems that can be in an equilibrium and while in equilibrium it holds that the number of pending requests in $Q(object[i]): i \in [1,M]$ is bounded.

\Subsection{Job arrival rates}
We assume that the time between two consecutive arrivals of $job_i$ to $thread[n]$ is a random variable $I[i,n]$ (inter-arrival period), where $i \in [1, J]$ and $n \in [1, N]$. We define the {\em job arrival rate}, $\lambda_{i,n}$, in which $job_i$ arrives at the system that then places $job_i$ in a (first in, first out) queue, $Q(thread[n])$, where $\lambda_{i,n}$ is a positive real number. The inter-arrival period of $I[i,n]$
follows an exponential distribution, $Exp(\lambda_{i,n})$. 
Note that this is a common way to model arrivals, e.g.~\cite{conf/sigmod/ChoG00}. As soon as $thread[n]$ becomes available, the system assigns to $thread[n]$ the job that is in $Q(thread[n])$'s top.
This work focuses on systems that can be in an equilibrium for which the number of pending jobs in $Q(thread[n])$ is bounded.

\Subsection{Work cycles: demand, supply and release}
\label{ss:work_cycles}
The thread work cycle, $cycle(thread[n], job_i)$, refers to the events that occur during the period that starts when the system assigns $job_i$ to $thread[n]$ and ends immediately before the next assignment of any job to $thread[n]$.
It starts with the event $\sigma_i(thread[n])$ in which the system assigns $job_i$ to $thread[n]$. 
%
%
It also includes the events in which: $thread[n]$ {\em demands} (requests) access to $object[k]$, denoted by $\delta_i(thread[n]$, $object[k]))$, the system {\em supplies} (provides) access to $object[k]$, denoted by $\sigma_i(object[k]$, $thread[n])$, and the event in which $thread[n]$ releases 
$object[k]$
denoted by $\phi_{i}(object[k],thread[n])$.
For simplicity, we refer to the sequence of these release events $\Phi_i(thread[n])$ $=$ $(\phi_i(object[k_1]$, $thread[n])$, $\ldots$, $\phi_i(object[k_\ell]$, $thread[n]))$ as a single event and assume that $thread[n]$ releases all its acquired objects instantaneously and immediately after the operation time, $O_i$, which is a random variable. 
Immediately after the event $\Phi_i(thread[n])$, the thread work cycle starts a (possibly zero length) {\em idle period}, before the system assigns the next (and possibly different than the previous) job to $thread[n]$ so that the next work cycle begins. 

\begin{wrapfigure}{r}{0.265\textwidth} 
\includegraphics[width=0.28\textwidth]{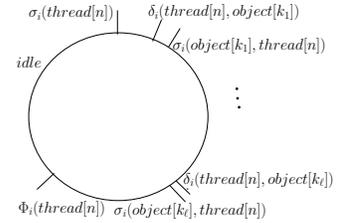}
%
\caption{The thread work cycle} 
\label{fig:work_cycle}
\end{wrapfigure}
We assume that events are instantaneous and mark them as points on a thread's work cycle (Figure \ref{fig:work_cycle}). 
Note however, that between a supply event and a demand event (as well as the last supply event and the release event), there is a random length period, i.e. the (random) acquisition period $A$ (and the operation time $O_i$, respectively), which refers to scheduling uncertainties. 
%
%
%
%
Hence, we denote $thread[n]$'s work cycle due to $job_i$ as
$cycle(thread[n], job_i)$ $\equiv$ $(\sigma_i(thread[n])$, $\delta_i(thread[n]$,  $object[k_1])$, $\sigma_i(object[k_1]$,
$thread[n])$, $\ldots$, $\delta_i(thread[n]$, $object[k_\ell])$, $\sigma_i(object[k_\ell]$, $thread[n])$, $\Phi_i(thread[n]))$.

\begin{wrapfigure}{r}{0.2\textwidth} 
\includegraphics[width=0.18\textwidth]{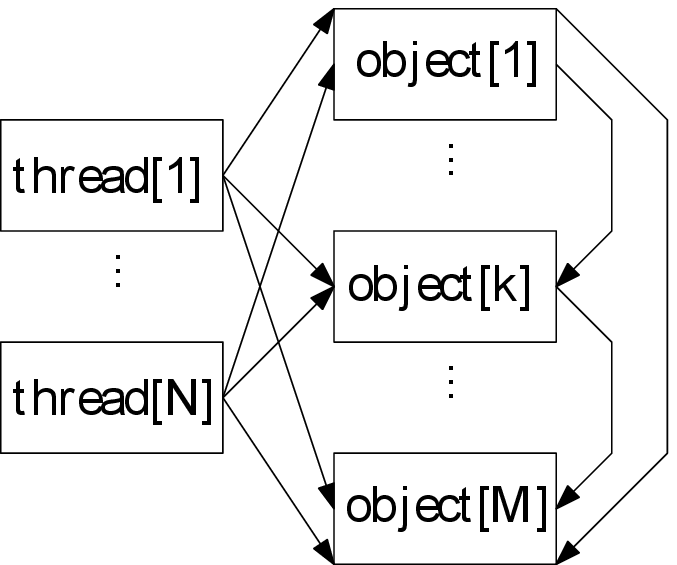}
\caption{An acquisition graph, $\dgraph$} 
\label{fig:routing_graph}
\end{wrapfigure}
\Subsection{Subpaths and acquisition graph}
%
%
Let $s$ (source) and $d$ (destination) be (possibly consecutive) items on a path. We define the set  $\epsilon(s,d)$ $=$ $\{(s$, $\bullet$, $d)$ $|$ $\exists job_i$ $:$ $path_i = ( \bullet$, $s$, $\bullet$, $d$, $\bullet)\}$ of all {\em subpaths} between $s$ and $d$, where  $\bullet$ denotes a finite, possibly empty, item sequence.
The {\em acquisition graph}, $\dgraph$ $=$ $(V,E)$, is a simple directed graph, where $V$ $=$ $\{ thread[1]$, $\ldots$, $thread[N]$, $object[1]$, $\ldots$, $object[M] \}$ are the system items (Figure~\ref{fig:routing_graph}).
The edges $E$ $=$ $\{ (s$,  $d)$ $|$ $\exists job_i$ $:$ $path_i$ $=$ $( \bullet$, $s$, $d$, $\bullet ) \}$ are two consecutive items on a path.

\newpage

\Subsection{Conditional and consecutive events}
\label{ss:cond-consec}
We consider an event that occurs at item $d$ (destination) in condition to an event occurrence at item $s$ (source), where $(\bullet,s,d,\bullet)$ is a subpath of $job_i$ and both events belong to the same work cycle of $job_i$ that $thread[n]$ carries out.

\Subsubsection{Conditional demand and supply events}
Denote by $\delta_{i, n}(d \,|\, s)$ $\equiv$ $\langle \delta_i(thread[n]$, $d)$ $|$ $\sigma_i(s$, $thread[n])\rangle$ the \emph{conditional (demand) event}, $\delta_i(thread[n]$, $d)$, in which $thread[n]$ requests access to object $d$ immediately after the supply event, $\sigma_i(s$, $thread[n])$. 
%
%
Note that the event $\sigma_i(s$, $thread[n])$ may refer to: (1) access to object $s$, or (2) $job_i$'s assignment to $s=thread[n]$ (Figure \ref{fig:work_cycle}), i.e., $\sigma_i(thread[n]) \equiv \sigma_i(thread[n],thread[n])$. 
%
%
E.g., $\delta_{i, n}(object[j]$ $|$ $object[k])$ denotes the conditional demand event $\langle \delta_i(thread[n]$, $object[j])$ $|$ $\sigma_i(object[k]$, $thread[n])\rangle$ in which $thread[n]$ requests access to $object[j]$ immediately after gaining access to $object[k]$, where $k \in [1,M-1]$, $j \in (k,M]$ and $n \in [1,N]$.  
Another example is the case $\delta_{i, n}(object[j]$ $|$ $thread[n])$, where the conditional demand event $\langle \delta_i(thread[n]$, $object[j])$ $|$ $\sigma_i(thread[n]$, $thread[n])\rangle$ refers to $thread[n]$'s request to access $job_i$'s first object, $object[j]$, immediately after the assignment of $job_i$ to $thread[n]$.  
In a similar manner, denote by $\phi_{i,n}(d|s)$ $\equiv$ $\langle \phi_i(d$, $thread[n])$ $|$ $\sigma_i(s$, $thread[n]) \rangle$ the \emph{conditional (release) event}, 
in which $thread[n]$ releases object $d$ at event $ \phi_i(d$, $thread[n])$ that occurs after the supply event, $\sigma_i(s$, $thread[n])$ and at the same work cycle 
(we will mainly use $\phi_{i,n}(s|s)$).

\Subsubsection{Events of arbitrary jobs and threads}
%
%
Sometimes we consider an arbitrary $job_i$ that an arbitrary $thread[n]$ carries out. 
We then write $\delta(d|s)$, $\sigma(s)$ and $\phi(d|s)$ instead of $\delta_{\bullet, \bullet}(d | s)$, $\sigma_{\bullet, \bullet}(s)$, and respectively, $\phi_{\bullet, \bullet}(d | s)$ when referring to events from the sets $\{ \delta_{i, n}(d | s) : i\in [1,J], n\in [1,N]\}$, $\{ \sigma_{i, n}(s) : i\in [1,J], n\in [1,N]\}$, and respectively, $\{ \phi_{i, n}(d | s) : i\in [1,J], n\in [1,N]\}$, where $i$ $\in$ $[1,J]$ and $n$ $\in$ $[1$, $N]$.
%
%
Given subpath $(\bullet$, $\ell_1$, $\ldots$, $\ell_k$, $\bullet)$, denote by $\delta(\ell_k| \ell_1, \ldots, \ell_{k-1})$ the occurrence of  $\delta(\ell_k|\ell_{k-1})$, which happens immediately after $\delta(\ell_{k-1}|\ell_{k-2})$, $\ldots$, $\delta(\ell_{2}|\ell_{1})$, where $k \in [1,M]$.

\Subsubsection{Consecutive events}
Let $i$, $j$ $\in$ $[1,J]$, $k$, $\ell$ $\in$ $[1,M]$, $n$, $n'$ $\in$ $[1,N]$, $\delta_{i,n}(object[k] | s)$  
%
%
and $\delta_{j,n'}(object[\ell] | s)$. 
%
%
%
We say that the event $\delta_{i,n}(object[k] \,|\, s)$ {\em occurs consecutively after} the event $\delta_{j,n'}(object[\ell] \,|\, s)$, when
$\delta_{j,n'}(object[\ell] \,|\, s)$ $=$ $\langle \delta_j(thread[n']$, $object[\ell]) \,|\, \sigma_j(s$, $thread[n'])\rangle$ is the first conditional demand event, that includes the supply event $\sigma_j(s$, $thread[n'])$, to occur after the conditional demand event $\delta_{i,n}(object[k]$ $|$ $s)$ $=$ $\langle \delta_i(thread[n]$, $object[k])$ $|$ $\sigma_i(s$, $thread[n])\rangle$, that includes the supply event $\sigma_i(s$, $thread[n])$.

%

\Subsection{Pairwise states and request probabilities}
The definition of the studied equilibria (at the system level) is based on item-level definitions that consider $\dgraph$'s edges, $(s,d) \in E$. 
We present a definition of the (pairwise) state, $c[s,d]$, which considers the delay, blocking and inter-demand periods that are related to the edge $(s,d)$ and its conditional events. These periods refer to the time it takes threads to request access to object $d$, and release it subsequently (after the acquisition of item $s$) as well as the time between such requests that are made by (possibly) different threads.
Moreover, when estimating the value of the pairwise-state, $c[s,d]$, we need to consider the probabilities that are related to the edge $(s,d)$ and its conditional events.

\begin{figure}[t!]
\begin{center}
    \includegraphics[width=0.45\textwidth]{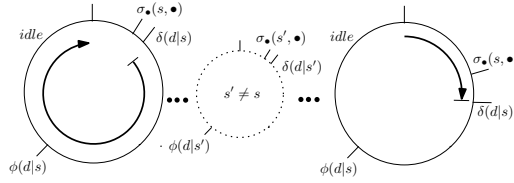}
\end{center}
\caption{$(s,d)$'s inter-demand period, $s.T[k]$, $d  =  object[k]$,
 i.e., the period between two consecutive $\delta(d | s)$ events.
}
\label{fig:sdidp}
\end{figure}

\begin{wrapfigure}{r}{0.175\textwidth} 
\includegraphics[width=0.2\textwidth]{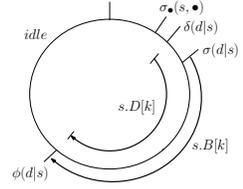}
%
\caption{$(s,d)$'s delay $s.D[k]$ and blocking $s.B[k]$ start from $\delta(d|s)$, and respectively, $\sigma(d|s)$, and both end at $\phi(d|s)$, $d=object[k]$} 
\label{fig:delay}
\end{wrapfigure}

\Subsubsection{Pairwise states}
%
%
We refer to {\em $(s, d)$'s pairwise state $c[s,d]$  $=$ $s.\langle T[k]$, $D[k]$, $B[k] \rangle$} as the tuple that includes the request completion rate, request delay, and respectively, blocking period with relation to the events $\delta(d|s)$, where $d = object[k]$ is an object and $k \in [1,M]$.
The {\em $(s,d)$'s request inter-demand period}, $s.T[k]$, refers to the period between the consecutive events, $\langle\delta_i(thread[n]$, $d)$ $|$ $\sigma_i(s$, $thread[n])\rangle$ and $\langle\delta_j(thread[n']$, $d)$ $|$ $\sigma_j(s$, $thread[n'])\rangle$ (Figure~\ref{fig:sdidp}), where $i,j \in [1,J]$ and $n,n' \in [1,N]$.
Notice that we can estimate the throughput, $1/s.T[k]$, which is the number of $\delta(d|s)$ requests per time unit. Moreover, we can consider the expected throughput, $1/E(s.T[k])$, where $E(s.T[k])$ is the expected value of $s.T[k]$.
Furthermore, {\em $(s,d)$'s delay}, $s.D[k]$, refers to a period that starts on the event $\delta(d|s)$ in which a thread requests access to object $d$ (immediately after gaining access to item $s$) and ends upon the event $\phi(d|s)$ in which that thread releases object $d$ (during the same work cycle). 
%
%
In addition, $(s,d)$'s blocking, $s.B[k]$, is the fraction of $(s,d)$'s delay $s.D[k]$ in which a thread blocks other threads from gaining access to object $d$ , i.e., the period between the event $\langle \sigma_i(d,thread[n]) | \sigma_i(s,thread[n]) \rangle$ in which $thread[n]$ gains access to object $d$, and the event $\phi_i(d,thread[n])$ in which $thread[n]$ releases $d$ (Figure~\ref{fig:delay}).
%
%
Note that each of $c[s,d]$'s three elements is a random variable (for which maintaining the first three moments provides sufficient accuracy). 


\Subsubsection{Pairwise request probabilities}
\label{sss:req_prob_def}
When estimating the pairwise-state $c[s,d]$, we use the {\em (pairwise) request probability}, $R(s,d)$, of events that are related to $\dgraph$'s edge $(s,d)$ to occur. When given the history of system events, we define the probability of the conditional demand event $\delta(d|s)$ to occur immediacy after the supply event $\sigma(s)$. We also consider the case in which the system does not have access to this information. 
In that case, we estimate $R(s,d)$ while assuming that $\delta(d|s)$ occurrence depends only on the system parameters, i.e., $N$, $M$, $\{job_i\}_{i\in [1,J]}$ and $\{\lambda_{i,n}\}_{i\in [1,J], n\in [1,N]}$, rather than requiring the availability of the event history.

For a randomly chosen work cycle that includes the event, $\sigma(s)$, of a thread gaining access to item $s$, we define $\Omega(s) = \{\delta(d|s) : (s,d) \in E \text{ is an edge in } \dgraph \} \cup \{\phi(s|s)\}$ as the probability space of the possible events to occur immediately after $\sigma(s)$. Moreover, $R(s,d)$ and $R(s,s)$ are the probabilities of $\Omega(s)$'s events $\delta(d|s)$, and respectively, $\phi(s|s)$.
Namely, $R(s,d)$ denotes the probability of a demand event, $\delta_\bullet(thread[\bullet], d)$ to occur immediately after the supply event, $\sigma_\bullet(s, thread[\bullet])$ and in the same (randomly chosen) work cycle.
%
%
Moreover, $R(s,s)$ denotes the probability of a release event, $\phi_\bullet(s, thread[\bullet])$, to occur immediately after its related supply event, $\sigma_\bullet(s$, $thread[\bullet])$ and during the same (randomly chosen) work cycle. Note that $R(s,s)$'s definition requires that $s$ is the last object for the thread to gain access to during that work cycle.
%

Since $R(s,d)$ and $R(s,s)$ depend on the history of events,
%
%
we further detail their definitions by using the notations $R_{t}(s,d)$ and  $R_{t}(s,s)$. We restrict the (random) choice of the work cycle to the time interval $t=[t_{start},t_{end}]$ and assume the awareness of all events that occurred in the system during that period.
For the time interval $t=[t_{start},t_{end}]$,
we define $R_{t}(s,d) = \alpha_{t}(s,d) / \eta_{t}(s)$, to be the number of $\delta(d|s)$ occurrences, over the number of $\sigma(s)$ occurrences and 
$R_{t}(s,s) = \beta_{t}(s)/ \eta_{t}(s)$ to be the number of $\phi(s|s)$ occurrences, over the number of $\sigma(s)$ occurrences, where $\#_{t}X$ denotes the number of event $X$ occurrences in $t=[t_{start},t_{end}]$. Moreover, $\alpha_{t}(s,d)$ $=$ $\Sigma_{i,n} \#_{t} \delta_{i,n}(d|s)$, when $s=object[j] : j \in [1,M-1]$, and $\alpha_{t}(s,d)$ $=$ $\Sigma_{i} \#_{t} \delta_{i,n}(d|s)$, when $s=thread[n] : n \in [1,N]$, the number of $\delta(d|s)$ events that occurred during $t=[t_{start},t_{end}]$. Furthermore, $\beta_{t}(s) = \Sigma_{i,n} \#_{t} \phi_{i,n}(s|s)$ and $\eta_{t}(s) = \Sigma_{i,n} \#_{t} \sigma_i(s, thread[n])$ denote the number of $\phi(s|s)$, and respectively, $\sigma(s)$ events that occurred during $t={[t_{start},t_{end}]}$. 
Note that profiling tools can be the basis for estimating $R_{t}(s,d)$ and $R_{t}(s,s)$. We also propose an estimation of $R(s,d)$ and $R(s,d)$ (Section~\ref{s:RoutingBlocking}) for the case in which these probabilities depend only on $N$, $M$, $\{job_i\}_{i\in [1,J]}$ and $\{\lambda_{i,n}\}_{i\in [1,J], n\in [1,N]}$ (and thus $t$'s history of events is not required to be available).

We define the {\em request probability matrix} $R$ to be a $(N+M)\times (N+M)$ row stochastic matrix. 
The matrix $R$ has a block form, where 
 $R_{N,N}$ is an $N\times N$ zero matrix, 
$R_{N,M}$ $=$ $(R(thread[n]$, $object[j]))_{n\in [1,N], j\in [1,J]}$ is an $(N\times M)$ matrix, 
$R_{M,N}$ is an $M \times N$ zero matrix and 
$R_{M,M}$ $=$ $(R(object[j]$, $object[k]))_{j,k \in [1,M]}$ is an (upper triangular) $M\times M$ matrix, i.e.,
\begin{equation}
\label{eq:R}
R = \left(\begin{array}{c|c}R_{N,N} & R_{N,M} \\ \hline R_{M,N} & R_{M,M}\end{array}\right)
\end{equation}

\Subsection{Item inter-demand period}
%
%
We refer to $object[k]$'s {\em inter-demand period}, $\sT_{object[k]}$, as the period  between two consecutive conditional (demand) events for accessing an object in $\{object[\ell]: \ell \in (k,M)\}$ 
immediately after gaining access to item $object[k]$ by two, possibly different, threads, where $k\in [1,M-1]$.
Namely, $\sT_{object[k]}$ is the period between $\delta(object[j]|object[k])$ and the successive conditional event $\delta(object[j']|object[k])$, where $j,j'$ $\in$ $(k$, $M]$.
We refer to $thread[n]$'s {\em inter-demand period},
%
%
$\sT_{thread[n]}$ as the period between $\delta_i(thread[n]$, $object[i_1])$ and the demand event $\delta_{i'}(thread[n]$, $object[i'_1])$, where $i,i'$ $\in$ $[1,J]$ and $object[i_1]$, $object[i'_1]$ are the first objects in the object vectors of $job_i$, and respectively, $job_{i'}$, where $thread[n]$ carries out $job_{i}$ and $job_{i'}$ consecutively. 
For example, we can use $1/E(\sT_{thread[n]})$ for estimating the number of job completions of $thread[n]$ per time unit, where $E(\sT_s)$ is $\sT_s$'s expected value.
In case $s$ $=$ $object[d]$, $1/E(\sT_{object[d]})$ estimates the number of job completions that include $object[d]$ per time unit.

We refer to $\kappa_{i,n}$ $=$ $c_{i,n} / \sT_{thread[n]}$ as $job_i$'s {\em completion rate} on $thread[n]$, which is the fraction of $thread[n]$'s inter-demand period, $\sT_{thread[n]}$, due to $job_i$, where $i\in [1,J]$, $n \in [1,N]$. We use $c_{i,n}$ as a constant for which $\forall n \in [1,N]:\Sigma_{i=1}^J$ $c_{i,n}$ $=$ $1$, and the weights $c_{i,n}$ are a function of the job arrival rates to $thread[n]$, for example $c_{i,n} = \lambda_{i,n}$ $/$ $(\Sigma_{i=1}^J \lambda_{i,n})$.


\Subsection{Shared-object system equilibria}
For a given system, $\psi({\dgraph})$ $=$ $\{c[s,d]\}_{(s,d) \in E}$ is the {\em system  state (set)}, where $\dgraph = (V,E)$ is the acquisition graph. For a given $\psi({\dgraph})$, the set $\tau(\dgraph) = \{\sT_{item}\}_{item \in V\setminus\{object[M]\}}$ is the {\em inter-demand period of the system}.

Suppose that a system is in a state in which the job arrival rates are equal to the job completion rates, i.e., $\forall i$ $\in$ $[1,J]$ $\forall n$ $\in$ $[1,N]$, $\lambda_{i,n}$ $=$ $\kappa_{i,n}$. We say that $\psi^\ast({\dgraph})$ $=$ $\{c^\ast[s,$ $d]\}_{(s,d) \in E}$, is the {\em shared-Object System Equilibrium} (OSE). Given $\psi^\ast({\dgraph})$, the respective inter-demand period of the system is $\tau^\ast({\dgraph})$ $=$ $\{\sT^\ast_{item}\}_{item\in V\setminus \{object[M]\}}$.
For a given $\varepsilon > 0$ and an OSE $\psi^\ast(\dgraph)$, we say that the system state $\psi(\dgraph)$ is an {\em $\varepsilon$-OSE} when $\forall\, item \in V\setminus\{object[M]\}, \sT^\ast_{item} \in \tau^\ast(\dgraph), \sT_{item} \in \tau(\dgraph) : |\sT^\ast_{item} - \sT_{item}| < \varepsilon$. Namely, the corresponding values of each item in $\tau(\dgraph)$ and $\tau^\ast(\dgraph)$ differ by less than $\varepsilon$.

Note that a system cannot always reach a state that satisfies the OSE conditions, and therefore an $\varepsilon$-OSE. Equilibria are {\em unreachable} when there is an item with a blocking period that is longer than (or equal to) the inter-arrival time of demand events to that item. For example, when the inter-arrival time of object requests is less or equal to the blocking period of that object. Note that in that case, the item's queue is increasing continuously.

\Section{The Solution Outline}
\label{s:probDesc}
%
%
We consider the case in which the job arrival rates can become equal to the job completion rates. We study how the system satisfies the OSE conditions both in exact and approximated manners.
%
%
We propose a procedure for finding the approximated equilibria, i.e., $\varepsilon$-OSEs. 
%
%
%
%
%
%
This procedure considers $\dgraph$'s paths, $(\bullet,s,d,\bullet)$, where $s$ is a system item, $d = object[k]$ and $k \in [1,M]$.

%
\begin{figure*}[t]
\begin{framed}

\noindent {\bf Estimating $R(s,d)$ (Section~\ref{s:RoutingBlocking})}
We define the following characteristic functions with values in $\{0,1\}$:
$starts_i(\langle s_1$, $\ldots$, $s_\ell \rangle)$ $\Leftrightarrow$ $objs_i$ $=$ $(s_1$, $\ldots$, $s_\ell$, $\bullet)$,
$includes_i(\langle s_1$, $\ldots$, $s_\ell \rangle)$ $\Leftrightarrow$ $objs_i$ $=$ $(\bullet, s_1$, $\ldots$, $s_\ell$, $\bullet)$ and
$ends_i(\langle s_1$, $\ldots$, $s_\ell \rangle)$ $\Leftrightarrow$ $objs_i$ $=$ $(\bullet$, $s_1$, $\ldots$, $s_\ell)$,
where $\langle s_1,\ldots, s_\ell \rangle$ is a vector of objects, $objs_i$ is the object vector of $job_i$ and $i\in [1,J]$.
When $s=thread[n]$ $:$ $n\in [1,N]$ and $k\in \{1,2\}$, 
$({\Sigma_{i} \lambda_{i,n} \cdot starts_i(\langle k\rangle)})/({\Sigma_{i} \lambda_{i,n}})$, and respectively, $({\Sigma_{i} \lambda_{i,n} \cdot includes_i(\langle 1,2\rangle)})({\Sigma_{i} \lambda_{i,n} \cdot includes_i(\langle 1\rangle)})$ 
approximate $R(s,d)$. Moreover, $\forall k \in \{1,2\}$ $:$ $({\Sigma_{i,n} \lambda_{i,n} \cdot ends_i(\langle k \rangle)})({\Sigma_{i,n} \lambda_{i,n} \cdot includes_i(\langle k \rangle)})$ approximates $R(s,s)$.

%
%

\noindent {\bf Estimating $s.B[k]$ (Section~\ref{s:blocking})}
Let $s$ $\in$ $\{thread[n]$, $object[1]\}$. Observe that the blocking periods $s.B[2]$ are only due to paths that finish in $object[2]$ and thus there are no dependencies for their estimation, i.e., $s.B[2]$ $=$ $A$ $+$ $R(s$, $object[2])$ $\cdot$ $R(object[2],object[2])$ $\cdot$ $f_{s,object[2]}$.
On the contrary, the blocking time of a thread's demand to $object[1]$, say $thread[n].B[1]$, depends on the possibility of $thread[n]$ to demand $object[2]$ and the respective delay, $object[1].D[2]$.
That is, $thread[n].B[1]$ $=$ $A$ $+$ $R(thread[n]$, $object[1])$ $\cdot$ $R(object[1]$, $object[1])$ $\cdot$ $f_{thread[n], object[1]}$ $+$ $R(thread[n]$, $object[1])$ $\cdot$ $R(object[1]$, $object[2])$ $\cdot$ $object[1].D[2]$, where  $f_{s, d}$ is the average of job operation times for jobs with paths $(\bullet,s,d)$ weighed by the probability, $R(s,d) \cdot R(d,d)$, of such events to occur. 

\noindent {\bf Estimating $\sT_{d}$ (Section~\ref{s:throughput})}
Let $arrivals(d)$ be the set of items $s$, such that a conditional demand event $\delta(d|s)$ can occur. Object $d$'s inter-demand period  is $\sT_d$ $=$ $\Sigma_{s\in arrivals(d)}(\omega(s$, $d)\cdot (\sT_s$ $+$ $A$ $+$ $R(s,d)$ $\cdot$ $R(d,d)$ $\cdot$ $f_{s,d}))/\Sigma_{s\in arrivals(d)}$ $\omega(s$, $d)$, where the weight is $\omega(s$, $object[k])$ $=$ $R(s$, $object[k])\cdot s.T[k]$, $arrivals(object[1]) = thread$, while $arrivals(object[2]) = thread \cup \{ object[1] \}$ and $thread$ $=$ $(thread[1]$, $\ldots$, $thread[N])$.

\end{framed}
\caption{Estimating $R(s,d)$, $s.B[k]$ and $\sT_{d}$ when $M=2$. }
\label{fig:M3Example}
\end{figure*}

%

\Subsection{Estimating $c[s,d]$ and $R(s,d)$}
We illustrate a solution for the case of $M=2$ objects and $N$ threads (Figure~\ref{fig:M3Example}) and outline the general case solution. 

\Subsubsection{The pairwise request probabilities}
The pairwise state, $c[s,d]$,  
and request probability, $R(s,d)$, are related to the conditional events, $\delta(d|s)$. 
When estimating the value of the pairwise state, we first need to estimate the probability for $\delta(d|s)$ to occur. 
Our approach considers both the case in which $R(s, d)$ and $R(s,s)$ are given and the case in which they depend only on the system parameters (Section~\ref{sss:req_prob_def}), i.e., $N$, $M$, $\{ job_i \}_{i \in [1,J]}$ and $\{\lambda_{i,n}\}_{i\in [1,J], n \in [1,N]}$.
Using the latter assumption, we estimate $R(s,d)$ by the sum of job arrival rates for which threads demand access to $d$ after the supply of item $s$ divided by the sum of job arrival rates due to which supply events for item $s$ occur (Figure~\ref{fig:M3Example}). 
Moreover, we estimate the probability $R(s,s)$ by the sum of job arrival rates such that their object vectors finish with $s$ divided by the sum of job arrival rates due to which supply events for item $s$ occur.

\Subsubsection{The thread's blocking periods}
Given the pairwise request probabilities, $R$, we estimate the blocking period, $s.B[k]$ (Section~\ref{s:blocking}), where $d = object[k]$. Item $d$'s blocking period depends in a recursive manner on the delay, cf. $thread[n].B[1]$'s dependency on $object[1].D[2]$ in Figure~\ref{fig:M3Example}, which in the general case appears as $s.B[k]$'s dependency on $object[k].D[k']:k' \in (k,M]$ in Lemma~\ref{lemma:blocking}. 
%
%
This dependency considers the delay that the thread experiences when further acquiring the remaining objects $t$ on the path $(\bullet,s,d,t,\bullet)$, which the thread needs for completing its job, where $t = object[\ell]$ and $\ell \in (k,M]$. 
Moreover, $s.B[k]$ depends on the {\em job completion period}, $f_{s,d}$, due to jobs with paths $(\bullet,s, d)$.
We refer to $(s,d)$'s (job) completion period, $f_{s,d}$, as the time between the event $\sigma(d|s)$, in which a thread gains access to object $d$ (after acquiring item $s$), which is the job's last object, and the event $\phi(d|s)$, in which that thread, immediately after executing the job operation, releases all objects (including $d$) that it had acquired during the work cycle that includes this two events (e.g., $f_{item[i_{k-1}],item[i_k]}$ is the period between $\sigma_{i_k}$ and  $\Phi_{i}$, see Figure~\ref{fig:work_cycle}). 
We detail the exact way in which such {\em forward dependencies} exist while the system satisfies the OSE condition (Section~\ref{s:blocking}). 
%

\Subsubsection{The item inter-demand period}
An item-level balance also exists and it is similar to the one that the system keeps for its threads (when satisfying the OSE conditions). Namely, the incoming rate of requests (demands) to access object $d$, has to balance with the inter-demand period, $\sT_d$, which is the time between two consecutive $\delta(x|d)$ and $\delta(y|d)$ events (where $x$ and $y$ are two, possibly different, objects). Note that $\sT_d$ depends on the rate of requests (demands) to access $d$ due to jobs with $(\bullet$, $d$, $d'$, $\bullet)$ paths, as well as, $(\bullet$, $d)$ paths. Thus, we estimate $\sT_d$, as the sum of the inter-demand period of every item $s$, $\sT_s$, and the job completion period $A$ $+$ $R(s,d)$ $\cdot$ $R(d,d)$ $\cdot$ $f_{s,d}$, times a weight that depends on the pairwise inter-demand period $s.T[k]$ plus the request probability $R(s,d)$, for every item $s$, such that $(s,d=object[k])$ is an edge in $\dgraph$. Note that $d$'s inter-demand period, $\sT_d$, has a {\em backward dependency}, i.e., it depends on the inter-demand periods $s.T[k]$ and $\sT_s$, where $(s,d)$ is an edge in $\dgraph$ and $d=object[k]$. We show the exact manner in which the system maintains this balance in Figure~\ref{fig:M3Example} for the case of $M=2$ and in Section~\ref{s:throughput} for the general case.

\Subsection{Resolving interdependencies}
Thus far, we have noticed several dependencies. 
For example, there are forward dependencies in which $(s,d)$'s blocking period, $s.B[k]$, depends on $t$'s delay, where $d = object[k]$, $t = object[\ell]$ and $\ell \in (k,M]$. 
Moreover, there is a backward dependency in which $(s,d)$'s inter-demand period, $\sT_d$, depends on the summation of pairwise inter-demand period, $s.T[k]$, for any path $(\bullet,s,d,\bullet)$. 
Note that more dependencies exist. The definition of the pairwise state $c[s,d]$, implies, for example, that $(s,d)$'s inter-demand period, $s.T[k]$, depends on $s$'s inter-demand period. Moreover, $d$'s delay depends on its blocking, $s.B[k]$, and the
$(s,d)$'s inter-demand period, $s.T[k]$.
Note that these pairwise state variables are inter-dependent due to blocking. 
We show a way to resolve these interdependencies by representing the thread work cycles as a subsystem in a way that is not subject to blocking and yet preserves the interdependencies that are related to the paths $(\bullet,s,d,\bullet)$ (Section~\ref{s:cont_graphs}). 
This approach for resolving forward and backward interdependencies is the basis of the proposed procedure for finding approximate equilibria.

\Subsection{Finding approximate equilibria}
\label{ss:finding_approx_ose}
We compute an approximated equilibrium, $\varepsilon$-OSE, when such is reachable. 
%
We propose a procedure that always halts (Algorithm~\ref{alg:sketch} presents the solution sketch and we detail the entire procedure in Section~\ref{s:appendix:ose}). 
It returns the system in an $\varepsilon$-OSE state whenever the job arrival and completion rates become equal, or indicates that the system cannot be in a state of an OSE.

The procedure starts with a system state that represents the case in which all queues are empty (line~\ref{alg:sketch:allEmpty}). It then estimates the state of a system in which threads can block one another, and the delay grows as more requests are pending in the queues.
The procedure works in iterations and decides when to stop using the system inter-demand period, $\{\sT_{item}\}_{item\in V \setminus \{ object[M] \}}$, i.e., it stops whenever there is no $item\in V \setminus \{ object[M] \}$ for which the change in $\sT_{item}$ is greater than $\varepsilon$ since the previous iteration (lines~\ref{alg:sketch:prevSet} to~\ref{alg:sketch:convergence}).


The procedure repeatedly improves an $\varepsilon$-OSE estimation until the system state satisfies the conditions of an approximated equilibrium. 
It deals with interdependencies using alternating backward and forward iterations (lines~\ref{alg:sketch:backward}, and respectively,~\ref{alg:sketch:forward}). Namely, we resolve the forward dependencies in which $(s$, $object[k])$'s blocking period, $s.B[k]$, depends on $object[\ell]$'s delay by {\em iterating backwards}, where $\ell \in (k,M]$. This backward iteration starts from $k=M$ and counts downwards. Moreover, it can estimate $s.B[k]$ 
(in a system that its state satisfies the equilibrium conditions), 
because all of $(s,object[k])$'s forward dependencies can be resolved. Similarly, we can use {\em forward iterations} for resolving backward dependencies with respect to $d$'s item inter-demand period, $\sT_{object[k]}$, because all of $object[k]$'s backward dependencies are resolved.

This loop also updates the thread inter-demand periods, i.e., the time between job completions (line~\ref{c:VisitedPathUpj}), and exits when no item's inter-demand period changes by at least $\varepsilon$ between every two iterations (line~\ref{alg:sketch:convergence}). 
Together with the estimation of $\sT_{thread[n]}$, the procedure checks whether the OSE condition is violated (Section~\ref{ss:tidp}), i.e., if the arrival rate, $\Sigma_{i=1}^J \lambda_{i,n}$, of jobs to $thread[n]$, is greater or equal than $1/blocking(n)$, where $blocking(n)$ $=$ $A$ $+$ $\Sigma_{k=1}^M R(thread[n]$, $object[k])$ $\cdot$ $thread[n].D[k]$ is the average time it takes to complete a job for $thread[n]$, and $1/blocking(n)$ is respective rate for $blocking(n)$. In case the OSE condition is violated, the loop breaks and the procedure returns.
Each iteration takes $O(M^2\cdot N^4 + M^3)$ time (see Section~\ref{s:appendix:alg_compl}).
%
%

\begin{algorithm}[t!]
\begin{small}
\caption{Finding an $\varepsilon$-OSE (procedure sketch)}
\label{alg:sketch}

{\bf Input:} $M$, $N$, $\{job_i\}_{i\in [1,J]}$, $\{\lambda_{i,n}\}_{n \in [1,N], i \in [1,J]}$,  $R$\;

{\bf Output:} $(thread[1,N], object[1,M-1], \{\sT_v\}_{v: v\neq object[M]})$\;

{\bf Macro: $\mathscr{S}(k) = \{thread\}\cup \{\{object[\ell]\} | \ell \in [1,k-1] \}$}\;

\noindent Start by supposing that all queues are empty\nllabel{alg:sketch:allEmpty}\;

\Repeat{the system has reached equilibrium (test for $\varepsilon$-OSE using the inter-demand period of every item)\label{alg:sketch:convergence}}{
\noindent {\bf let} $prevSet \gets$ item inter-demand periods\nllabel{alg:sketch:prevSet}\;
\For{$k = M$~{\bf to}~$1$ $(*$ backward iteration $*)$\label{alg:sketch:backward}}{\ForEach{$\forall S_k \in \mathscr{S}(k), \forall s\in S_k$}{Estimate $s.T[k]$, $s.D[k]$ by resolving the $(s, object[k])$-subsystem (Section~\ref{s:cont_graphs})\nllabel{alg:sketch:subI}\;
Estimate $s.B[k]$ (Section~\ref{s:blocking})\nllabel{alg:sketch:EstimateB}\;  }}
\lForEach{$n \in [1,N]$}{Estimate $\sT_{thread[n]}$ and upon OSE condition violation call {\bf return}(`no OSE')\nllabel{c:VisitedPathUpj}}
\For{$k = 1$~{\bf to}~$M-1$ $(*$ forward iteration $*)$ \label{alg:sketch:forward} }{\ForEach{$\forall S_k \in \mathscr{S}(k), \forall s\in S_k$}{Estimate $s.T[k]$, $s.D[k]$ by resolving the $(s, object[k])$-subsystem (Section~\ref{s:cont_graphs})\; \nllabel{alg:sketch:subII}}
Estimate $\sT_{object[k]}$ (Section~\ref{s:throughput})\;\nllabel{alg:sketch:EstimateT}}}
\Return{$(thread[1,N], object[1,M-1], \{\sT_v\}_{v: v\neq object[M]})$}\label{alg:sketch:until}\;
\end{small}
\end{algorithm}

\section{Background knowledge}
\label{s:bckgr}
Our solution uses tools from queueing networks~\cite{bolch2006queueing}.
Although queueing theory celebrated results provide closed forms for single queues, e.g., M/M/c, M/G/1~\cite{adan2002queueing}, and queueing networks, e.g., BCMP~\cite{baskett1975open}, Gordon-Newell~\cite{gordon1967closed}, closed form results are far from been the common case. Specifically, there are no relevant closed-form results that can be used for systems like ours in which a thread can block other threads for a non-exponentially distributed period. 
%
%
%
%
Ramesh and Perros~\cite{DBLP:journals/pe/RameshP01} consider a message passing system of multi-tier server networks in which processes communicate iteratively via what is known in the system community as synchronous I/O (and sometimes called blocking I/O).
%
%
Our solution requires resolving interdependencies. We use the thread work cycle for showing that our subsystems (Section~\ref{s:cont_graphs}) can represent these interdependencies. We then show that Ramesh-Perros subsystems~\cite{DBLP:journals/pe/RameshP01} can analyze our subsystems and resolve their interdependencies iteratively. We find $\varepsilon$-OSEs in a similar manner. 
Namely, we use a framework proposed by Baynat and Dallery~\cite{DBLP:journals/pe/BaynatD96} 
%
%
for estimating the system state, in a similar manner to Ramesh-Perros~\cite{DBLP:journals/pe/RameshP01}.
%
%
The authors of~\cite{DBLP:journals/pe/RameshP01,DBLP:journals/pe/BaynatD96} demonstrate the convergence of their iterative methods via numerical experiments. Baynat and Dallery~\cite{DBLP:journals/pe/BaynatD96} show that each iteration has polynomial running time, 
%
which is $O(M \cdot N^4)$ for the OSE case (Lemma~\ref{lem:framework_complexity}).

In the remainder of this section we present the stochastic process through which we calculate the probability of a thread to be idle, using the job arrival process and the time that it takes a thread to complete an arbitrary job (Section~\ref{s:appendix:jobAssignmentProbability}), as well as, a version of the Baynat-Dallery framework adapted to shared-object systems (Section~\ref{s:appendix:framework_details}).

\Subsection{Idle thread probability}
\label{s:appendix:jobAssignmentProbability}
We analyze the stochastic process in which jobs arrive in a shared-object system and are assigned to a thread, $thread[n]$, whenever a previously assigned job is completed (departure) or the thread is idle.
This process is characterized by the inter-arrival time of jobs to $thread[n]$, $I_n$ and the respective blocking period $B_n$, during which $thread[n]$ carries out the job.
We explain how an existing method can help us to obtain the probability of $thread[n]$ to be idle, $u_n$ (unoccupied). 

We define a Markov chain that has the structure of a quasi-birth-death process (QBD)~\cite{feldman2010applied}, using $I_n$ and $B_n$.
We consider general distributions of arrival $I_n$ and departure $B_n$ stochastic processes, to which we match Coxian-$2$ distributions, using their first three moments~\cite{tayfur1997performance}.
We consider (Section~\ref{s:preliminaries}) the arrival process of $job_i$ to $thread[n]$ to follow an exponential distribution $I[i,n]$ $\sim$ $Exp(\lambda_{i,n})$, as in~\cite{conf/sigmod/ChoG00}.
Therefore, we can assume that $I_n$ follow an exponential distribution with parameter $\Sigma_{i=1}^J \lambda_{i,n}$, where $J$ is the number of the system's jobs (as in~\cite{feldman2010applied}).
Moreover, we obtain the first three moments of $B_n$ by the moments of the $thread[n]$'s inter-demand period, $\sT_{thread[n]}$, that we calculate in Section~\ref{s:throughput}.
By applying moment matching to the first three moments of $I_n$ and $B_n$, we obtain the arrival, and respectively, departure rates of the continuous-time Markov chain.
Thus, we can define a quasi-birth-death (QBD) process that considers the growth and decrease in the number of pending jobs to be assigned to $thread[n]$, where each state determines the arrival and departure state of the respective Coxian-2 distribution.

We apply the Matrix Geometric Method (MGM) \cite{DBLP:books/daglib/0018540} to find the steady-state probabilities for each state of $thread[n]$, due to the Markov chain's QBD structure.
We then obtain $u_n$ directly since it equals the sum of the QBD process' probabilities for having no pending jobs, i.e., $u_n = u_{n,1} + u_{n,2}$, where $u_{i,n}$ is the steady-state probability of having no pending jobs and the arrival of a new job being while on phase $i$ of the Coxian-2 distribution, $i\in \{1,2\}$.
The MGM is an iterative method and its running time is in $O(I_{MGM}\cdot m^3)$, where $m$ is the maximum number of states in each QBD level and $I_{MGM}$ is the number of iterations until the method converges.
Since for our purposes, $m$ is a constant, e.g., $m=4$ in the modeling described here, $I_{MGM}$ is a small number and the method converges quadratically, we assume that the running time of the MGM is (practically) constant.
We base our assumption that $I_{MGM}$ is a small number on an example that Latouche and Ramaswami~\cite{latouche1993logarithmic} give.
They argue that in practical settings, $I_{MGM}$ is a small number, considering the case of $I_{MGM}>40$.
Here, $J>10^{12}$, where $J$ is the number of jobs.

\Subsection{Baynat-Dallery framework}
\label{s:appendix:framework_details}

%
As a background knowledge, we discuss a variation on Baynat and Dallery's framework~\cite{DBLP:journals/pe/BaynatD96} that we adapt to the context of shared-object systems (Algorithm~\ref{alg:framework}).
%
%
The $BDF()$ function denotes our adapted version of Baynat and Dallery's framework. For every $s\in S_k$, this function takes a contention subsystem, $\cs(S_k, k)$, which is the tuple $(\ch(S_k, k)$, $(\sR_s)_{s\in S_k}$, $(\sB_s)_{s\in S_k})$, as an input and returns an estimation of the delay $s.D[k]$ and inter-demand period $s.T[k]$.
Namely, $(s.T[k]$, $s.D[k])_{s \in S_k}$ $=$ $BDF(\cs(S_k, k))$.
The solution of Baynat and Dallery is based on iterative approximations of the demand arrival rates and request completion rate to $object[k]$ with the ones in the subgraph $\ch_s(S_k, k)$ $=$ $(\sV_s, \sE_s)$, and vice versa, until, for every $s\in S_k$, their absolute difference is below a given threshold.

We complete this section with a detailed explanation of Algorithm~\ref{alg:framework}.
The procedure starts by
an initialization phase (lines~\ref{alg:framework:init_start}--\ref{alg:framework:init_end}), 
which is followed by a repeat-until loop (lines~\ref{alg:framework:repeat_start}--\ref{alg:framework:repeat_end}) and
the output calculation (lines~\ref{alg:framework:output_start}--\ref{alg:framework:output_end}) before returning the output (line~\ref{alg:framework:return_output}).

\Subsubsection{Variables}
%
%
Lemma~\ref{lem:cs_representation} of Appendix~\ref{s:appendix:proof} shows that a contention subsystem (Section~\ref{ss:cont_subsys_def}) represents the dependencies among the threads in a shared-object system with respect to its state.
Let $s\in S_k$ denote a thread, if $S_k = thread$, or an object, if $S_k = \{object[j]\}$, where $j\in [1,k-1]$.
For item $v\in \sV_s$, we define $\sI_s(v)$, $\sB_s(v)$ and $\sC_s(v)$ to be item $v$'s the inter-arrival time $\sI_s(v)$, blocking period $\sB_s(v)$, and respectively, inter-demand period $\sC_s(v)$.
With respect to the $BDF()$ function, $\sI_s(v)$ is the time between two demand events to item $v$, $\sB_s(v)$ is the blocking period of an arbitrary demand event to $v$, and respectively, $\sC_s(v)$ is the time between two release events on $v$.
Note that when $s$ is a thread ($S_k = thread$), $\sI_s(s)$ is the time between two object release events by $s$, $\sB_s(s)$ is the time from an object release event until the next job completion by $s$, and respectively, $\sC_s(s)$ is the time between two job completions by $s$.
Baynat and Dallery approximate $\sI_s(v)$, $\sB_s(v)$ and $\sC_s(v)$ using exponential distributions with parameters (rates) $\gamma_s(v)$, $\mu_s(v)$, and respectively, $\nu_s(v)$.
Note that these random variables depend only on $s$ and $v$, due to the definition of the contention subsystem (Section~\ref{ss:cont_subsys_def}) and Lemma~\ref{lem:cs_representation} of Appendix~\ref{s:appendix:proof}.
The $BDF()$ function estimates $\gamma_s(v)$, $\mu_s(v)$ and $\nu_s(v)$ and uses them to compute $s.D[k]$ and $s.T[k]$ for every $s\in S_k$.

We define the subgraph $\ch_s(S_k, k)$ $=$ $(\sV_s, \sE_s)$ of $\ch(S_k, k)$ (Section~\ref{ss:cont_subsys_def}).
Note that when $S_k$ $=$ $thread$, thread $s$ $=$ $thread[n]$ requests to access only to items in $\ch_s(S_k,k)$'s subgraph.
Similarly, when $S_k$ $=$ $\{object[\ell]\}$, a thread that has access to $s$ $=$ $object[\ell]$ requests to access only to $\ch_s(S_k,k)$'s items, where $\ell \in [1,k-1]$.
%
%
Baynat and Dallery treat these subgraphs as Gordon-Newell networks~\cite{gordon1967closed, buzen1973computational}.
The $BDF()$'s repeat-until loop (lines~\ref{alg:framework:repeat_start}--\ref{alg:framework:repeat_end}) alternates between computing $\gamma_s(v)$, $\mu_s(v)$ and $\nu_s(v)$ for the Gordon-Newell network defined by the subgraph $\ch_s(S_k,k)$, for every $s\in S_k$ and the same values for every individual item $v\in \sV$.
The iterations stop when for two consecutive loops there is no $s\in S_k$ and $v\in \sV_s$, such that the values of $\mu_s(v)$ differ more than a given $\varepsilon$ (line~\ref{alg:framework:repeat_end}).

\Subsubsection{Initialization phase}
The $BDF()$ function initializes $\mu_s(v)$ with $1/E(\sB_s(v))$ (lines~\ref{alg:framework:init_start}--\ref{alg:BDF:init_mu}) and computes the steady-state probabilities of a thread to demand or have access to item $v$, after the supply of item $s$.
It does that through the function $stationary()$, which takes the stochastic matrix $\sR_s$ as an input. The function $stationary()$ outputs the steady state vector, $steadyStateProbabilities_s$, which has the size of $|\sV_s|$. Moreover, $steadyStateProbabilities_s$ satisfies the equations $\pi \cdot \sR_s = \pi$ and $\Sigma_{i=1}^{|\sV_s|}\pi_i = 1$ (line~\ref{alg:framework:init_end}).

\Subsubsection{The repeat-until loop}
The $BDF()$ function's repeat-until loop calculates $\gamma_s(v)$ (lines~\ref{alg:framework:subgraph_start}--\ref{alg:framework:set_lambda}), $\nu_s(v)$ (lines~\ref{alg:framework:item_for}--\ref{alg:framework:item_end}) and $\mu_s(v)$ (lines~\ref{alg:framework:new_blocking_start}--\ref{alg:framework:new_blocking_end}).
We calculate the Gordon-Newell normalizing constant (line~\ref{alg:framework:GN}) and the marginal probabilities of Gordon-Newell (lines \ref{alg:framework:subgraph_for}--\ref{alg:framework:marg_probs1}).
Using these marginal probabilities, we calculate $\gamma_s(v)$ for every $s\in S_k$ and $v\in \sV_s$ (line~\ref{alg:framework:set_lambda}).

We find the marginal probability of an item to be idle through the $idleProb()$ function (lines~\ref{alg:framework:item_for}--\ref{alg:framework:item_marg_probs1}) and then calculate $\nu_s(v)$ (line~\ref{alg:framework:item_end}) for every $s \in Z(v)$, where $Z(v)$ $=$ $\{s \in S_k~|$ $v \in \sV_s\}$ (line~\ref{alg:BDF:theta}).
The $idleProb()$ function calculates item $v$'s marginal probability to be idle through the underlying Markov chain of a multi-class queue with exponential arrivals ($\gamma_s(v)$ for every $s \in Z(v)$). It also calculates the blocking periods ($\mu_s(v)$ for every $s\in Z(v)$). Note that the queue length is limited by the maximum number of pending demands ($N$ when $S_k = thread$ and 1 when $S_k = \{object[j]\}$, $j\in [1,k-1]$)~\cite{DBLP:journals/pe/BaynatD96,bolch2006queueing}.

The calculation $\mu_s(v)$'s new estimates (lines~\ref{alg:framework:new_blocking_start}--\ref{alg:framework:new_blocking_end}) happens before the next iteration. In order to check the convergence condition, each iteration
begins with storing in the $oldValues$ variable the last estimations of $\mu_s(v)$ for every $s\in S_k$ and $v\in \sV_s$ (line~\ref{alg:framework:repeat_end}). 

\Subsubsection{$BDF()$'s output}
We estimate $s.D[k]$ through the delay in $object[k]$'s queue in the contention subsystem (line~\ref{alg:framework:output_end}).
Moreover, we obtain an estimation of the inter-demand period $s.T[k]$ through $1/\gamma_s(object[k])$ (line~\ref{alg:framework:output_end}).
The $BDF()$ function returns the output in line~\ref{alg:framework:return_output}.


\begin{algorithm*}[th!]
\caption{The $BDF()$ function for estimating delay and pairwise inter-demand period through a contention subsystem}
\label{alg:framework}


{\bf Input:} $\cs(S_k$, $k)$ $=$ $(\ch(S_k$, $k)$, $(\sR_s)_{s\in S_k}$, $(\sB_s)_{s\in S_k})$\label{alg:BDF:input}\;
{\bf Output:} $(s.T[k]$, $s.D[k])_{s \in S_k}$\label{alg:BDF:output}\;

{\bf Macros:}\\
$converged(prev,curr) =$ {$(\nexists\, \mu_s(v) \in prev, \mu'_s(v) \in curr$ $:$ $|\mu_s(v) - \mu'_s(v)|\geq\varepsilon)$}\label{alg:BDF:convergence}\;
$Z(v) = \{s \in S_k | v \in \sV_s\}$\label{alg:BDF:theta}\;

\Begin{

\ForEach{$s \in S_k$\label{alg:framework:init_start}}{\lForEach{$v\in \sV_s$}{$\mu_s(v) \gets 1/E(\sB_s(v))$}\label{alg:BDF:init_mu}}

\lForEach{$s \in S_k$}{$(steadyStateProbabilities_s(v))_{v\in \sV_s} \gets stationary(\sR_s)$\label{alg:framework:init_end}}

\Repeat{$converged(oldValues,\{\mu_s(v)\}_{s \in S_k, v \in \sV})$\label{alg:framework:repeat_end}}{\label{alg:framework:repeat_start}
{\bf let} $oldValues = (\mu_s(v))_{s\in S_k, v\in V}$\label{alg:framework:old_values}\;

\ForEach{$s \in S_k$\label{alg:framework:subgraph_start}}{
$GordonNewellConstant_s \gets \Sigma_{v \in \Theta(s)} steadyStateProbabilities_s(v)/\mu_s(v)$\label{alg:framework:GN}\;
\ForEach{$v \in \sV_s$\label{alg:framework:subgraph_for}}{
$subgraphMarginalProbs_{s,v}(1) \gets steadyStateProbabilities_s(v)/(\mu_s(v) \cdot GordonNewellConstant_s)$\label{alg:framework:marg_probs0}\;
$subgraphMarginalProbs_{s,v}(0) \gets 1 - subgraphMarginalProbs_{s,v}(1)$\label{alg:framework:marg_probs1}\;
$\gamma_s(v) \gets \mu_s(v)\cdot \left(subgraphMarginalProbs_{s,v}(1)/subgraphMarginalProbs_{s,v}(0)\right)$\label{alg:framework:set_lambda}\;
}
}

\ForEach{$v \in V$\label{alg:framework:item_for}}{
\ForEach{$s\in Z(v)$\label{alg:framework:item_start}}{
$itemMarginalProbs_{s,v}(0) \gets idleProb(v,s, \{\gamma_s(v)\}_{s \in S_k}, \{\mu_s(v)\}_{s \in S_k}, (\sR_s)_{s\in S_k})$\label{alg:framework:item_marg_probs0}\;
$itemMarginalProbs_{s,v}(1) \gets 1 - itemMarginalProbs_{s,v}(0)$\label{alg:framework:item_marg_probs1}\;
}
\lForEach{$s\in Z(v)$}{$\nu_s(v) \gets \gamma_s(v)(itemMarginalProbs_{s,v}(0)/itemMarginalProbs_{s,v}(1))$\label{alg:framework:item_end}}
}

\ForEach{$s \in S_k$\label{alg:framework:new_blocking_start}}{
\lForEach{$v \in \sV_s$}{$\mu_s(v) \gets \nu_s(v)$\label{alg:framework:new_blocking_end}}
}

}


\ForEach{$s\in S_k$\label{alg:framework:output_start}}{
{\bf let} $(s.D[k], s.T[k]) = (1/(\mu_s(object[k]) \cdot subgraphMarginalProbs_{s,object[k]}(0)),1/\gamma_s(object[k])$\label{alg:framework:output_end}\;}
\Return{$(s.T[k], s.D[k])_{s\in S_k}$\label{alg:framework:return_output}\;}
}

\end{algorithm*}

\Section{Request probabilities}
\label{s:RoutingBlocking}
%
%
%
%
The fact that the pairwise request probabilities depend on the arrival rates of the corresponding jobs is the basis of our estimation of $R(s,d)$ (Lemma~\ref{lem:req_probs}) as in the network approximations~\cite{feldman2010applied}, where $(s,d)$ is an edge in $\dgraph$. 
%
In Lemma~\ref{lemma:prob} we prove that our estimations of $R(s,d)$ and $R(s,s)$ define indeed a probability, i.e., for any item $s$, $R(s,s) + \Sigma_{d\neq s} R(s,d) = 1$.
%
This implies that the probability matrix $R$, which contains the estimates of $R(s,d)$ and $R(s,s)$, is a stochastic matrix.
%
%
\begin{lemma}
\label{lem:req_probs}
Equation~\ref{eq:Rsdthread} and Equation~\ref{eq:Rsdobject} 
approximate $R(s$, $d)$, 
when $s=thread[n]$ $:$ $n\in [1,N]$, and respectively, $s=object[j]$ $:$ $j\in [1,M-1]$ , see Figure~\ref{fig:M3Example} for the definitions of the characteristic functions $starts()$, $includes()$ and $ends()$.
Moreover, for any object $s$, Equation~\ref{eq:Rssobject} approximates $R(s,s)$.
\end{lemma}
\begin{small}
\begin{equation}
\label{eq:Rsdthread}
R(thread[n], d) \approx 
\frac{\left(\Sigma_{i} \lambda_{i,n} \cdot starts_i(\langle d\rangle)\right)}{\left(\Sigma_{i} \lambda_{i,n}\right)}
\end{equation}
\end{small}
\begin{small}
\begin{equation}
\label{eq:Rsdobject}
R(object[j], d) \approx 
\frac{\Sigma_{i} \lambda_{i,n} \cdot includes_i(\langle object[j],d\rangle)}{\Sigma_{i} \lambda_{i,n} \cdot includes_i(\langle object[j]\rangle)}
\end{equation}
\end{small}
\begin{small}
\begin{equation}
\label{eq:Rssobject}
R(s,s) \approx 
\frac{\left(\Sigma_{i,n} \lambda_{i,n} \!\cdot\! ends_i(\langle s\rangle)\right)}{\left(\Sigma_{i,n} \lambda_{i,n} \!\cdot\! includes_i(\langle s\rangle)\right)}
\end{equation}
\end{small}
Equation~\ref{eq:Rsdthread} estimates $R(thread[n]$, $d)$ by the sum of arrival rates of jobs that start with object $d$ to $thread[n]$, divided by the sum of the arrival rates of all the jobs that are assigned to $thread[n]$.
Equation~\ref{eq:Rsdobject} estimates $R(s$, $d)$ by the sum of arrival rates of jobs that include the subvector $(s,d)$ in their object vector (to any thread), divided by the sum of arrival rates of jobs that include the subvector $(s)$ in their object vector (to any thread), where $s=object[j]$ and $j\in [1,M-1]$.
Equation~\ref{eq:Rssobject} estimates $R(s$, $s)$ by the sum of arrival rates of jobs that their object vector ends with the subvector $(s)$ (to any thread), divided by the sum of arrival rates of jobs that include the subvector $(s)$ in their object vector (to any thread), where $s=object[j]$ and $j\in[1,M]$.
Note that in any other case, we define $R(s,d)=0$, since no conditional demand events $\delta(d|s)$ or conditional release events $\phi(s|s)$ occur in these cases.

%
\label{s:appendix:proofs}

We prove that the estimation of the request probabilities in Lemma~\ref{lem:req_probs} also defines a probability (Lemma~\ref{lemma:prob}). 

\begin{lemma}
\label{lemma:prob}
For $R(s,d)$'s estimation (Lemma~\ref{lem:req_probs}), it holds that: 
(1) $\Sigma_d R(thread[n],d)$ $=$ $1$, where $n\in [1,N]$ and $d$ is an object,
(2) $R(s,s)$ $+$ $\Sigma_{d\neq s} R(s$, $d)$ $=$ $1$, and
(3) $R$ is a row stochastic matrix.
\end{lemma}

\proof{
Equation~\ref{eq:SigmadRthread} demonstrates claim (1). 
%
\begin{eqnarray}
\label{eq:SigmadRthread}
%
\Sigma_d R(thread[n], d)& = & \frac{\Sigma_d\Sigma_{i} \lambda_{n,i} \cdot starts_i(\langle d\rangle)}{\Sigma_{i} \lambda_{n,i}}\\
& = & \frac{\Sigma_{i} \lambda_{n,i} \Sigma_d starts_i(\langle d\rangle)}{\Sigma_{i} \lambda_{n,i}} \nonumber \\
& = & \frac{\Sigma_{i} \lambda_{n,i} \cdot 1}{\Sigma_{i} \lambda_{n,i}} \nonumber \\
& = & 1 \nonumber 
\end{eqnarray}
%
For any object $s\neq object[M]$, Equation~\ref{eq:Sigmaddneqs} demonstrates claim (2) due to Equation~\ref{eq:SigmaSimple}, which holds since a job object vector that includes object $s$, either ends with $s$ or includes more items $d\neq s$. 
\begin{small}
\begin{eqnarray}
\label{eq:Sigmaddneqs}
%
& & R(s,s) + \Sigma_{d\neq s} R(s, d)  \\
& = & \frac{\Sigma_{n,i} \lambda_{n,i} \cdot ends_i(\langle s\rangle)}{\Sigma_{n,i} \lambda_{n,i} \cdot includes_i(\langle s\rangle)} 
+ 
\frac{\Sigma_{d\neq s}\Sigma_{n,i} \lambda_{n,i} \cdot includes_i(\langle s,d\rangle)}{\Sigma_{n,i} \lambda_{n,i} \cdot includes_i(\langle s\rangle)}  \nonumber \\
&& \nonumber \\
& = &
\frac{\Sigma_{n,i} \lambda_{n,i} \cdot ends_i(\langle s\rangle) + \Sigma_{d\neq s}\Sigma_{n,i} \lambda_{n,i} \cdot includes_i(\langle s,d\rangle)}{\Sigma_{n,i} \lambda_{n,i} \cdot includes_i(\langle s\rangle)}  \nonumber \\
&& \nonumber \\
& = &
\frac{\Sigma_{n,i} \lambda_{n,i} \left(ends_i(\langle s\rangle) + \Sigma_{d\neq s}includes_i(\langle s,d\rangle)\right)}{\Sigma_{n,i} \lambda_{n,i} \cdot includes_i(\langle s\rangle)}  \nonumber \\
&& \nonumber \\
& = &
\frac{\Sigma_{n,i} \lambda_{n,i} \cdot includes_i(\langle s\rangle)}{\Sigma_{n,i} \lambda_{n,i} \cdot includes_i(\langle s\rangle)}  \nonumber \\
&& \nonumber \\
&=&1 \nonumber 
\end{eqnarray}
\end{small}
%
%
\begin{equation}
\label{eq:SigmaSimple}
%
ends_i(\langle s\rangle) + \Sigma_{d\neq s}includes_i(\langle s,d\rangle) = includes_i(\langle s \rangle)
\end{equation}

As for claim (3), we note that claims (1) and (2) imply that our estimation of the block matrices $R_{N,M}$ and $R_{M,M}$ of the matrix $R$ are row stochastic (Equation~\ref{eq:R}). 
Since the block matrices $R_{N,N}$ and $R_{M,N}$ are zero matrices, our estimation of $R$ forms a row stochastic matrix.
}

\Section{Blocking Periods}
\label{s:blocking} 
We estimate $(s,d)$'s the blocking period, $s.B[k]$, using the request probabilities and the job completion periods, where $d=object[k]$, $k\in [1,M]$ and $(s,d) \in E$ is an edge in $\dgraph$.
This blocking period is an effect of multiple threads' job paths, i.e., the $(\bullet,s,d)$ and the remaining $(\bullet,s,\bullet,d,\bullet)$ paths.
The former case corresponds to the job completion period (Lemma~\ref{lemma:job_completion}), whereas the latter depends on the delay of acquiring the path remaining objects (Lemma~\ref{lemma:blocking}).

\Subsubsection{Job Completion Periods}
\label{ss:job_compl_per}
%
%
%
%
%
%
In the case of $(\bullet,s,d,\bullet)$ paths, the period for acquiring the remaining objects varies according to $(s,d)$'s delay, the job completion period depends only on known distributions (the job operation times), and the probability that the related events occur.
%
However, for the case of $(\bullet,s,d)$ paths, $f_{s,d}$ is the operation time average due to jobs with $(\bullet,s,d)$ paths, weighted by the probability for the related events to occur (Lemma~\ref{lemma:job_completion}). This allows us to estimate $(s,d)$'s blocking time, $s.B[k]$.
Lemma~\ref{lemma:job_completion} 
%
%
defines 
%
%
%
%
$f_{s,d}$ $=$ $\Sigma_{i=1}^J$ $weight_i(s,d)$ $\cdot$ $O_i$ as a weighted average of all the operation times, 
where $weight_i(s,d)$ $=$ $(\Sigma_{n=1}^N$ $\lambda_{i,n}$ $\cdot$ $ends_i(\langle s,d\rangle)$ $/$ $W_{s,d}$, if $s$ is an object, 
$weight_i(s,d)$ $=$ $\lambda_{i,n}$ $\cdot$ $ends_i(\langle s,d\rangle)$ $/$ $W_{s,d}$, if $s = thread[n]$
and $W_{s,d}$ $=$ $\Sigma_{i=1}^J weight_i(s,d)$ is the sum of all weights.
Note that the weights depend on the arrival rates of jobs with object vectors that end with $(s,d)$, if $s$ is an object and $(d)$, if $s$ is a thread.

\begin{lemma} 
\label{lemma:job_completion}
$(s,d)$'s completion period is a weighted average, $f_{s,d} = \Sigma_{i=1}^J weight_i(s,d) \cdot O_i$, of the respective job operation times.
\end{lemma}

\Subsubsection{Acquiring the remaining objects}
We give an example of how to estimate $s.B[k]$ in a system with two objects (Figure~\ref{fig:M3Example}).
%
%
%
For the general case of $M$ objects, we also have to account for $(\bullet, s$, $\bullet$, $r$, $d, \bullet)$ paths, as in the proof of Lemma~\ref{lemma:blocking}.

\begin{lemma}
\label{lemma:blocking}
%
%
The $(s,d)$'s blocking period can be estimated by $s.B[k]$ $=$ $A$ $+$ $R(s$, $object[k])$ $\cdot$ $R(object[k]$, $object[k])$ $\cdot$ $f_{s,object[k]}$ $+$ $\Sigma_{k' = k+1}^M (\Pr(\rho = \rho(k'))$ $\cdot$ $object[k].D[k'])$, where $\rho = (\bullet$, $s$, $\bullet$, $object[k]$, $\bullet)$ is a path that includes items $s$ and $d=object[k]$, as well as $\Pr(\rho = \rho(k'))$ is the probability of having $\rho(k') = (\bullet$, $s$, $\bullet$, $object[k]$, $object[k']$, $\bullet)$ as a path.
\end{lemma}

\proof{
Let $d$ $=$ $object[k]$ and $d'$ $=$ $object[k']$.
We define the probability $\Pr(\rho$ $=$ $(\bullet$, $s$, $\bullet$, $d$, $d'$, $\bullet))$ $=$ $R(s$, $d)$ $\cdot$ $R(d$, $d')$ $+$ $\Sigma_{r=\tilde{s} + 1}^{k-1} R(s$, $object[r])$ $\cdot$ $[\Sigma_{j = 1}^{k-r} R^j(object[r]$, $d)]$ $\cdot$ $R(d$, $d')$ of a path $\rho$, 
%
%
where $\tilde{s}$ $=$ $0$, if $s$ $=$ $thread[n]$, $\tilde{s}$ $=$ $j$, if $s$ $=$ $object[j]$. $R^j(\alpha$, $\beta)$ is the probability that the path $(\alpha,\bullet,\beta)$ includes $j-1$ intermediate items between $\alpha$ and $\beta$, or equivalently, the probability of an $\dgraph$'s path from $\alpha$ to $\beta$ to include $j$ edges, which is given by the $(\alpha,\beta)$ element of the $j$-th power of the stochastic matrix $R$.
}
Lemma~\ref{lemma:blocking}'s proof details $s.B[k]$'s forward dependency via its recursive equation that estimates by computing $s.B[M]$ blocking periods, before $s.B[M-1]$ and so on. Moreover, it depends on the delays, $object[k].D[k']$, for acquiring the remaining job objects. We use backward iterations (Section~\ref{s:probDesc}) to resolve these dependencies (Section~\ref{s:cont_graphs}).

\Section{Item inter-demand periods}
\label{s:throughput}
The item inter-demand period allows us to decide on $\varepsilon$-OSE's condition satisfaction.
We estimate the item inter-demand period, which together with the blocking period, $s.B[k]$ (Section~\ref{s:blocking}), are essential for calculating the delay, $s.D[k]$ (Section~\ref{s:cont_graphs}), where $d=object[k]$, $k\in [1,M]$ and $(s,d) \in E$ is an edge in $\dgraph$.

\Subsubsection{Object inter-demand period}
\label{ss:object_inter_demand_periods}
Recall that item $d$'s inter-demand period, $\sT_d$, is the time between two demand completions on item $d$, which in equilibrium equals the time between two demand arrivals to $d$.
Therefore, our inter-demand period estimation, $\sT_{d}$, for an object $d$ depends on $(s,d)$'s inter-demand periods, which are the time intervals between two consecutive demands to $d$ that first acquire item $s$, the inter-demand period $s.T[k]$, and the period during which the threads complete such demands.

The estimation of object $d$'s inter-demand period, $\sT_{d}$, is the average of $s$'s inter-demand periods plus the time that demands for $(\bullet,s,d)$ paths block $d$, where $(s,d)$ is an edge in $\dgraph$. This time of demands is $A+R(s,d)\cdot R(d,d) \cdot f_{s, d}$, weighted by the probability of $(\bullet$, $s$, $d$, $\bullet)$ paths and the $(s,d)$'s inter-demand period, $s.T[k]$, of demands to $d$ after the supply of item $s$, due to $(\bullet, s, d, \bullet)$ paths (the weights are normalized by their sum).
Namely, the average sums the period between two demand completions of an item $s$ and the job completion time, and the weights incorporate the item dependencies through $s.T[k]$ for all $(\bullet,s,d, \bullet)$ paths and their probability to occur ($d=object[k]$).
By generalizing the example system with $M=2$ objects and $N$ threads (Figure~\ref{fig:M3Example}) for every $(s,d) \in E$, we obtain Lemma~\ref{lemma:agg_throughput}.

\begin{lemma}
\label{lemma:agg_throughput}
The inter-demand period of $d=object[k]$ is:
$\sT_d$ $=$ $(\Sigma_{s \in arrivals(d)}$ $\omega(s$, $d)\cdot (\sT_s$ $+$ $A$ $+$ $R(s,d)\cdot R(d,d) \cdot f_{s,d}))$ $/$ $(\Sigma_{s \in arrivals(d)}$ $\omega(s$, $d))$, where $arrivals(object[k])$ $=$ $thread$ $\cup$ $\{object[j]$ $|$ $j \in [1,k-1]\}$ and $\omega(s$, $object[k])=$ $R(s$, $object[k])\cdot s.T[k]$.
\end{lemma}

Lemma~\ref{lemma:agg_throughput} details $\sT_d$'s backward dependency via its recursive equation that calculates $\sT_{object[j]}$, $j \in [1,k-1]$ before calculating $\sT_{object[k]}$.
Moreover, the calculation of $\sT_{thread[n]}$ and $\sT_{object[k]}$ depends on the inter-demand periods, which is $thread[n].T[k]$, of $(thread[n], object[k])$ and respectively, $s.T[k]$, where $k\in [1,M]$ and $s\in thread$ $\cup$ $\{object[j]$ $|$ $j\in [1,k-1]\}$.
We use forward iterations (Section~\ref{s:probDesc}) for resolving such dependencies (Section~\ref{s:cont_graphs}).

\Subsubsection{Thread inter-demand period}
\label{ss:tidp}
We estimate $thread[n]$'s inter-demand period, $\sT_{thread[n]}$, by a Markov process that considers both the arrival process of jobs to $thread[n]$ and the time it takes $thread[n]$ to complete a job.

The way that we estimate $\sT_{thread[n]}$, uses the $augmntThreadBlock(I(n)$, $blocking(n))$ function.
The function input includes the inter-arrival times $I(n)$ and $blocking(n)$ $=$ $A$ $+$ $\Sigma_{k=1}^M R(thread[n]$, $object[k])$ $\cdot$ $thread[n].D[k]$, i.e., the average time it takes $thread[n]$ to complete a job. 
Moreover, $augmntThreadBlock(I(n)$, $blocking(n))$ outputs the estimation of $\sT_{thread[n]}$ and checks if the OSE condition is violated (line~\ref{c:VisitedPathUpj} in Algorithm~\ref{alg:sketch} of Section~\ref{ss:finding_approx_ose}), i.e., the rate $\Sigma_{i=1}^J \lambda_{i,n}$ that defines $I(n)$ (we assume $I(n)$ to be exponentially distributed, as in~\cite{conf/sigmod/ChoG00}) is greater or equal than $1/blocking(n)$. 
The function analyzes a queue using Queuing Theory~\cite{adan2002queueing}.
In our case, we characterize the queue by matching the first three moments of $I(n)$, and respectively, $blocking(n)$ to Coxian-2 distributions (Section~\ref{s:appendix:jobAssignmentProbability}).
%

%


\label{s:appendix:requestGenerationTime}


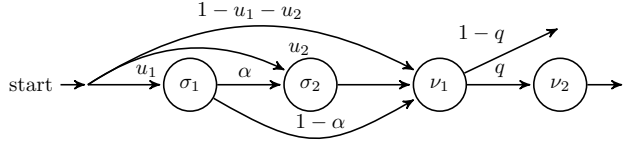
\begin{figure}[t!]
\begin{center}

\newcommand{\scaleOfG}{0.8}

\begin{tikzpicture}[->,>=stealth',shorten >=1pt,auto,node distance=2cm,
                    semithick,scale=\scaleOfG, every node/.style={scale=\scaleOfG}] 
  \tikzstyle{every state}=[draw]

\node[state]	(A)	{$\sigma_1$};
\coordinate[draw=none,left= 1cm of A,initial]          (initial);
\node[state]	(B)	[right of=A] {$\sigma_2$};
\node[state]	(C)	[right =1cm of B] {$\nu_1$};
\node[state]	(D)	[right of=C] {$\nu_2$};
\coordinate[draw=none,right=0.5cm of D]         (terminal1);
\coordinate[draw=none,above=0.4cm of D]          (terminal2);

\path (A) edge              node {$\alpha\,\,\,$} (B)
(B) edge              node {} (C)
(C) edge              node {$q$} (D)
(A) edge         [bend right, distance = 50]     node {$\,\,\,1-\alpha$} (C)
(initial)     edge          node {$\qquad u_1$} (A)
(initial)     edge          [bend left,right]    node {$\,\,\,\qquad\qquad u_2$} (B)
(initial)     edge          [bend left,distance=65]    node {$1- u_1- u_2$} (C)
(D)     edge     [scale=0.5cm]         node {} (terminal1)
(C)     edge              node {$1-q$} (terminal2);
\end{tikzpicture}

\end{center}
\caption{\small Request generation (Markov) process}
\label{fig:ast}

\end{figure}

%

Recall that $\sT_{thread[n]}$ denotes the period between two consecutive $\delta(d|thread[n])$ and $\delta(d'|thread[n])$ events.
The system assigns a job pending to a thread's queue, say $thread[n]$, whenever that thread becomes available, where $n\in [1,N]$.
After this assignment (and a random acquisition period of $A$), the event $\delta(d|thread[n])$ occurs in which the thread requests to access the job's first object; denoted by $d$.

%
%


We estimate $\sT_{thread[n]}$ by a Markov chain that depicts both $I(n)$ and $blocking(n)$ (as in Ramesh and Perros~\cite{DBLP:journals/pe/RameshP01} but with adaptation to shared-object systems).
In Figure~\ref{fig:ast}, $\sigma_1$, $\sigma_2$ and $\alpha$ define the Coxian-$2$ job arrival process at $thread[n]$ and $\nu_1$, $\nu_2$ and $q$ define the Coxian-$2$ of the $thread[n]$'s job completion period ($blocking(n)$), which we obtain by applying moment matching~\cite{tayfur1997performance} on the first three moments of $I(n)$ and $blocking(n)$.
We denote with $u_j$ the probability that upon an arbitrary job completion $thread[n]$ becomes idle and the job arrival process is in phase $j$ $\in$ $\{1,2\}$, as in~\cite{DBLP:journals/pe/RameshP01} 
%
(Section~\ref{s:appendix:jobAssignmentProbability}).
%
%
We define $\sT_{thread[n]}$ to follow a phase-type distribution (PH)~\cite{adan2002queueing}, which is defined using the initial probability vector, $c$ $=$ $( u_1$, $u_2$, $1- u_1- u_2$, $0)$, and a transition matrix $Q$ (Figure~\ref{fig:ast}), where $Q=\left[\begin{array}{c|c} S& S^0\\ \hline \mathbf{0}&0 \end{array}\right]$ in its block form, the $(x,y)$ element of $S$, $S(x,y)$, equals the rate in state $x$ times the transition probability to state $y$, $x,y \in \{ \sigma_1, \sigma_2, \nu_1, \nu_2 \}$, and $S^0 = -S\mathbf{1}$.
The $m$-th moment of $\sT_{thread[n]}$ is $E[\sT_{thread[n]}^{m}] = (-1)^{m}m! c S^{-m}\mathbf{1}$~\cite{adan2002queueing}.

\Section{Resolving Dependencies} 
\label{s:cont_graphs}
We showed how to estimate the blocking period on an object, $s.B[k]$, while depending on the delay for acquiring other objects, $object[k].D[k']$ (Section~\ref{s:blocking}), as well as how to estimate the inter-demand periods $\sT_{thread[n]}$ and $\sT_{object[k]}$, while depending on the $s.T[k]$ pairwise inter-demand periods, due to $(\bullet$, $s$, $object[k]$, $\bullet)$ paths (Section~\ref{s:throughput}).
Recall that these variables are inter-dependent due to blocking.
Theorem~\ref{thm:BaynatDallery} demonstrates that we can resolve these interdependencies by representing the thread work cycles as a subsystem in a way that is not subject to blocking and yet preserves these interdependencies. 
%
After the definition of the subsystem, we provide the key steps of the proof by looking into the case of $M=3$ (Section~\ref{ss:cont_graph_example}). 
We prove Theorem~\ref{thm:BaynatDallery}'s in Section~\ref{s:appendix:proof} using background knowledge~\cite{DBLP:journals/pe/RameshP01} (which we refer to in Section~\ref{s:bckgr}). 

\begin{theorem}
\label{thm:BaynatDallery}
Let $s\in S_k$ and $d=object[k]$, where $S_k \in \{thread\}$ $\cup$ $\{\{object[j]\}$ $|$ $j \in [1,k)\}$. The Baynat-Dallery framework can approximate the delay $s.D[k]$ and pairwise inter-demand period $s.T[k]$ through the contention subsystem $\cs(S_k$, $k)$ $=$ $(\ch(S_k$, $k)$, $(\sR_s)_{s\in S_k}$, $(\sB_s)_{s\in S_k})$, where $S_k \in \{thread\}\cup s_k$ and $s_k = \{\{object[i]\}$ $|$ $i \in [1,k-1]\}$. The running time of each framework iteration is $O(M \cdot N^4)$.
\end{theorem}


\Subsection{Contention subsystems} 
\label{ss:cont_subsys_def}
Given any pair of system items, $s$ and $d = object[k]$, 
we calculate the (pairwise) state $c[s, d]$ for $\dgraph$'s edge $(s$, $d)$ using a construction that we name {\em contention subsystem} $\cs(S_k,k)$, where $k \in [1,M]$, $s \in S_k$ and $S_k$ is either the set of all threads, $S_k = thread$, representing $(thread[n],d,\bullet)$ job paths, or a set including a single object, $S_k = \{object[j]\}$, $j\in [1,k-1]$, representing $(\bullet,object[j],d,\bullet)$ paths.
For every item $s \in S_k$, we use the thread work cycle, $cycle(thread[n]$, $job_j)$ of $thread[n]$, $n\in [1,N]$, carrying out $job_j$, (Section~\ref{s:preliminaries}), to show that the contention subsystem $\cs(S_k,k)$ represents the state of the shared-object system, with respect to the interdependencies among the delay $s.D[k]$ and the pairwise inter-demand period $s.T[k]$ when the blocking time is known. 
%

%

Let $object[k] \in \{object[1]$, $\ldots$, $object[M]\}$ as well as 
$s_k = \{ \{object[1]\}$, $\ldots, \{object[k-1]\}\}$ (i.e., $s_k=\emptyset$ when $k=1$) and
$S_k$ $\in$ $\{thread\}$ $\cup$ $s_k$, where $thread$ $=$ $\{thread[1]$, $\ldots$, $thread[N] \}$.
Moreover, let $Rel(thread$, $k)$ $=$ $[1$, $M]$ $\setminus$ $\{k\}$ and 
$Rel(\{object[i]\}$, $k)$ $=$ $[i+1,M]\setminus \{k\}$, where $object[i] \in s_k$.
We partition the $(\bullet$, $s$, $\bullet)$ paths to three sets,
$\sP(s, d)$ $=$ $\cup_{\ell\in [1,3]} \sP_\ell$, where
$\sP_1 =$ $\{path\,|\, path$ $=$ $(\bullet$, $s$, $d$, $\bullet)\}$, 
$\sP_2$ $=$ $\{ path \,|\, path$ $=$ $(\bullet$, $s$, $object[i]$, $\bullet$, $d$, $\bullet)$ $\land$ $i\in Rel(S_k$, $k)$ $\setminus$ $[k+1,M]\}$, 
$\sP_3$ $=$ $\{path \,|\, path$ $=$ $(\bullet$, $s$, $\bullet)$ $\land$ $d$ $\notin$ $path\}$ and $Rel(thread$, $k)$ $=$ $[1$, $M]$ $\setminus$ $\{k\}$.

A {\em contention subsystem}, is denoted by $\cs(S_k$, $k)$ $=$ $(\ch(S_k$, $k)$, $(\sR_s)_{s\in S_k}$, $(\sB_s)_{s\in S_k})$ and defined by (1)--(3).
%
%

(1) The {\em contention graph} 
$\ch(S_k$, $k)= (\sV, \sE)$ has the set of vertices $\sV$ $=$ $\cup_{s\in S_k}\sV_{s}$, 
and the set of edges $\sE$ $=$ $\cup_{s\in S_k}\sE_{s}$, such that for every $s \in S_k$, $\sV_{s}$ $=$ $\{s\}$ $\cup$ $\{d\}$ $\cup$ $\{relay(s,j)$ $|$ $j$ $\in$ $Rel(S_k$, $k)\}$ and $\sE_{s}$ $=$ $\sE^1_{s}$ $\cup$ $\sE^2_{s}$ $\cup$ $\sE^3_{s}$, 
where
$\sE^1_{s}$ $=$ $\{(s$, $d)$, $(d$, $s)\}$, 
$\sE^2_{s}$ $=$  $\cup_{j\in Rel(S_k,k)\setminus [k+1,M]}\{(s$, $relay(s$, $j))$, $(relay(s$, $j)$, $d)$, $(d$, $s)\}$, and 
$\sE^3_{s}$ $=$  $\cup_{j\in Rel(S_k,k)}\{(s$, $relay(s$, $j))$, $(relay(s$, $j)$, $s)\}$, i.e., $\sE^{k}_s$ corresponds to the partition $\sP_k$, where $k \in [1,3]$.
Note that $\ch(S_k$, $k)$ is a simple graph, i.e., there are no multiple edges between two vertices.
Moreover, $relay(s,j)$ is a distinct copy of a relay $object[j]$, $j\in Rel(S_k,l)$, for $s \in S_k$.

(2) The {\em request probability matrices} $\sR_s$ for $\ch(S_k$, $k)$ $=$ $(\sV$, $\sE)$ and $s \in S_k$.
$\sR_s(s$, $d)$ depicts the probability of a path $(\bullet$, $s$, $d$, $\bullet)$.
Moreover, $\sR_s(s$, $relay(s,j))$ depicts the probability of a path $r$ $=$ $(\bullet$, $s$, $object[j]$, $\bullet)$, while
$\sR_s(relay(s,j)$, $d)$ depicts the probability that $r$ $=$ $(\bullet$, $s$, $object[j]$, $\bullet$, $d$, $\bullet)$.
Furthermore, $\sR_s(v$, $s)$, $v\in \sV_s \setminus \{s\}$, depicts the probability of a thread becoming idle or starting a new job after the completion of a job, which is a certain event and therefore $\sR_s(v,s) = 1$, $v\in \sV_s \setminus \{s\}$.  


(3)
The {\em blocking periods}, $(\sB_s)_{s\in S_k}$, where $\sB_s$ is a function over the set of items in $\sV_s$, and $s\in S_k$ refers to the thread blocking periods on each of $\sV_s$'s items. 
Note that $E_s^i$ forms a directed circle in $\ch(S_k,k)$, where $s\in S_k$ and $i\in [1,3]$. 
A demand request to $d$ that follows the supply of $s \in S_k$ and possibly the supply of a relay object, $object[j]$, $j\in Rel(S_k,k) \setminus [k+1,M]$, is blocking $d$ for $\sB_s(d)$ $=$ $s.B[k]$ time.
A demand request to server $relay(s,j)$ that follows the supply of $s \in S_k$, blocks that server for a period of $\sB_s(relay(s$, $j))$ $=$ $s.D[j]$, if $j\in [k+1,M]$ and $\sB_s(relay(s$, $j))$ $=$ $W(s, relay(s,j), d)$, if $j\in Rel(S_k,k)\setminus [k+1,M]$,
where $W(s, relay(s, j), d)$ equals the delay $s.D[j]$ minus the blocking period of a possibly subsequent demand event to $d=object[k]$.
Once a job is completed, another demand event follows the supply of $s \in \sV_s$ after a period of $\sB_s(s) = \sT_s$.

\Subsection{The case of systems with $M=3$ objects} 
\label{ss:cont_graph_example}
%

\begin{figure*}[t!]
\begin{subfigure}[b]{0.24\textwidth}
\centering
\includegraphics[width=\textwidth]{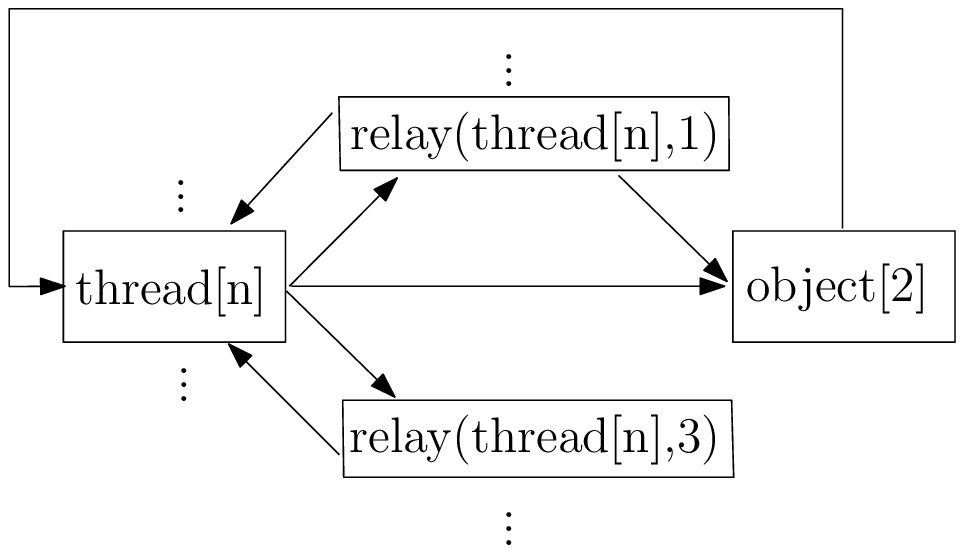}
\caption{Contention graph for $M=3$}
\label{fig:contsubsys_example:graph}
\end{subfigure}
\begin{subfigure}[b]{0.24\textwidth}
\centering
\includegraphics[width=\textwidth]{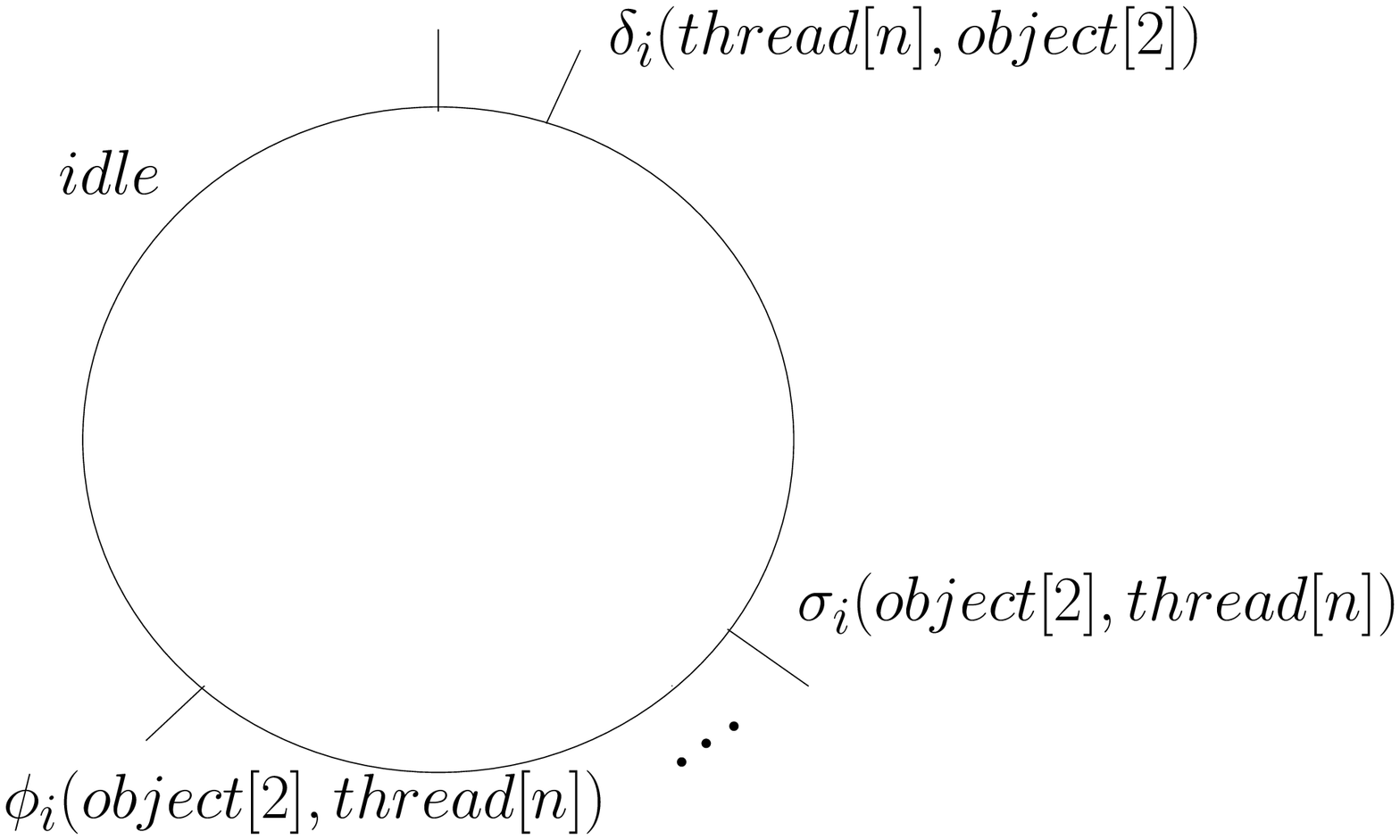}
\caption{The case of $\sP_1$}
\label{fig:contsubsys_example:case1}
\end{subfigure}
%
%
\begin{subfigure}[b]{0.24\textwidth}
\centering
\includegraphics[width=\textwidth]{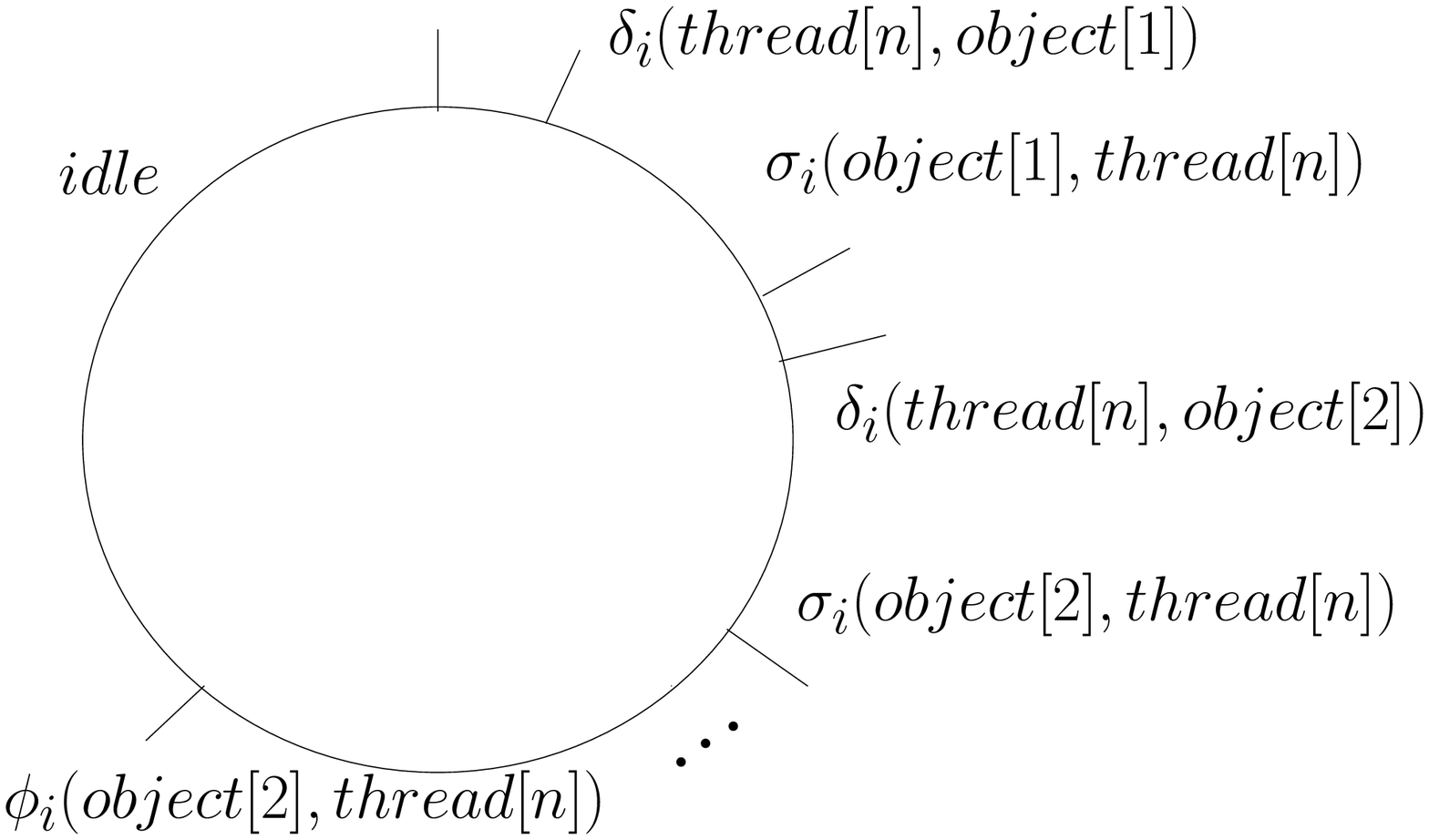}
\caption{The case of $\sP_2$}
\label{fig:contsubsys_example:case2}
\end{subfigure}
\begin{subfigure}[b]{0.24\textwidth}
\centering
\includegraphics[width=\textwidth]{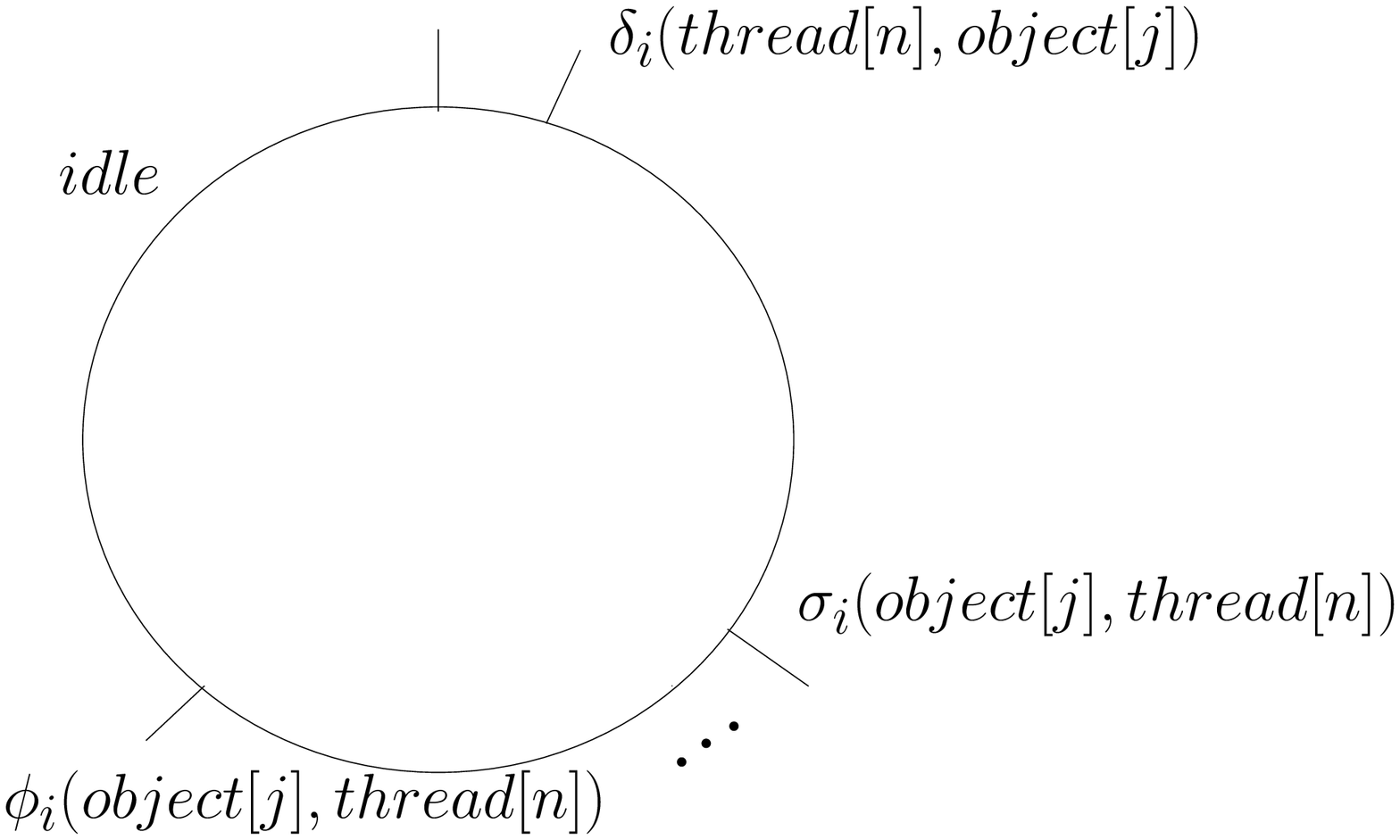}
\caption{The case of $\sP_3$}
\label{fig:contsubsys_example:case3}
\end{subfigure}
%
%
\caption{The contention graph for $\cs(thread, 2)$ and the work cycles partitions,
$\sP(s, d)$ $=$ $\cup_{\ell\in [1,3]} \sP_\ell$, of $(s$, $\bullet)$ paths,  where
$\sP_1 =$ $\{\chi$ $|$ $\chi$ $=$ $(s$, $d$, $\bullet)\}$, 
$\sP_2$ $=$ $\{ \chi$ $|$ $\chi$ $=$ $(s$, $object[1]$, $d$, $\bullet)\}$, $\sP_3$ $=$ $\{\chi$ $|$ $\chi$ $=$ $(s$, $\bullet)$ $\land$ $d$ $\notin$ $\chi\}$, $s=thread[n]$ and $d=object[2]$. 
}
%
%
%
\label{fig:contsubsys_example}
\end{figure*}

%
We use an illustrative example that shows how the contention subsystem of $S_k$ $=$ $thread$ and $d$ $=$ $object[2]$ represents the dependencies among the threads of a system with $M=3$ objects and $N$ threads, with respect to the delays and pairwise inter-demand periods when the blocking times and the item inter-demand periods are known. 
We construct the contention subsystem 
$\cs(thread$, $2)$ $=$ $(\ch(thread$, $2)$, $(\sR_s)_{s\in thread}$, $(\sB_s)_{s\in thread})$
based on the work cycles related to the delay $thread[n].D[2]$ and inter-demand period $thread[n].T[2]$, for every $s$ $=$ $thread[n]$ $\in$ $thread$ and $d = object[2]$.
We explain the representation of work cycles by a contention graph, which is illustrated in Figure~\ref{fig:contsubsys_example:graph}, and the adaptation of the request probabilities and blocking times to the ones of the contention subsystem.
The challenge here is to demonstrate that a dynamic system that is based on correlated events with dependencies that are due to blocking and follow non-deterministic schedules can be represented by these subsystems. 
After demonstrating this part of the proof, the rest of the proof follows by matching between the subsystems presented here to the one by Ramesh-Perros~\cite{DBLP:journals/pe/RameshP01}, which use a framework proposed by Baynat and Dallery~\cite{DBLP:journals/pe/BaynatD96} for estimating our system's state.


\Subsubsection{Contention graph of $\cs(thread, 2)$}
Let $\ch(thread$, $2) = (\sV, \sE)$ for $S_k = thread$ and $d=object[2]$. (Figure~\ref{fig:contsubsys_example:graph}).
Given an arbitrary thread, $s=thread[n]$, $n\in [1,N]$, let
$\sP(s$, $d)$ $=$ $\cup_{\ell\in [1,3]} \sP_\ell$ be a partition of $(s$, $\bullet)$ paths, where
$\sP_1 =$ $\{path$ $|$ $path$ $=$ $(s$, $d$, $\bullet)\}$, 
$\sP_2$ $=$ $\{ path$ $|$ $path$ $=$ $(s$, $object[1]$, $d$, $\bullet)\}$
and $\sP_3$ $=$ $\{path$ $|$ $path$ $=$ $(s$, $\bullet)$ $\land$ $d$ $\notin$ $path\}$.
%
%
Note that for brevity, we write $s=thread[n]$, $o_1=object[1]$, $o_3=object[3]$, and $d=object[2]$ throughout this example, but we clarify this notation when needed.
Moreover, let 
$\sV$ $=$ $\cup_{n\in [1,N]}\sV_{s}$ be the union of
$\sV_{s}$ $=$ $\{s$, $relay(s,1)$, $d$, $relay(s$, $3)\}$,
where $relay(s$, $j)$, $j\in \{1$, $3\}$, are $s$'s distinct copies of a relay object, $object[j]$,
which allow us to distinguish paths with respect to threads. 
The contention graph's nodes $s$, $relay(s,j)$ and $d=object[k]$ represent $s$, $object[j]$ and, respectively $d$, in the shared-object system, where $j\in \{1, 3\}$.


%
The edges $\sE$ $=$ $\cup_{n \in [1,N]} \sE_{s}$ follow the path partition cases, $\{\sE_\ell\}_{\ell\in [1,3]}$.
Let $job_i$ be a job that $s=thread[n]$ carries out. 
The edge sets $\sE_\ell$, where $\ell \in [1,3]$, are defined as follows.

$\bullet$ $\sP_1$'s case refers to work cycles, for which $s$ demands access to $d$, once it is assigned with $job_i$.
When $s$ gains access to $d$, $job_i$ might require $s$ to demand access to $o_3$. 
Upon $job_i$'s completion $s$ releases any acquired object. 
Thus, 
the edges in $\sE_1$ $=$ $\{(s$, $d)$, $(d$, $s)\}$ (Figure~\ref{fig:contsubsys_example:graph})
represent the work cycle subvectors
$(\delta_i(s$, $d))$, and respectively, $(\sigma_i(d$, $s)$, $\ldots$, $\phi_i(d$, $s)$, $\ldots)$ (Figure~\ref{fig:contsubsys_example:case1}).

$\bullet$ $\sP_2$'s case refers to work cycles, for which $s$'s demand for access to $o_1$ is followed by $s$'s demand for access to $d$ after $o_1$'s supply, which is then followed by $job_i$'s completion. 
Thus, the edges in the set $\sE_2$ $=$
$\{(s$, $relay(s$, $1)$, $(relay(s$, $1)$, $d)$, $(d$, $s)\}$ (Figure~\ref{fig:contsubsys_example:graph})
represent the work cycle subvectors
$(\delta_i(s$, $o_1))$, $(\sigma_i(o_1$, $s)$, $\delta_i(s$, $d))$, and respectively, $(\sigma_i(d$, $s)$, $\ldots$, $\phi_i(d$, $s)$, $\ldots)$ (Figure~\ref{fig:contsubsys_example:case2}). 

$\bullet$  $\sP_3$'s case refers to work cycles, for which $s$ demands access to $object[j]$ and then completes $job_i$, where $j\in \{1,3\}$. Therefore, 
the edges in $\sE_3 = \{(s, relay(s, j)), (relay(s, j), s)\}$ (Figure~\ref{fig:contsubsys_example:graph}),
represent the subvector
$(\delta_i(s, object[j]))$, and respectively, $(\sigma_i(object[j])$, $s)$, $\ldots$, $\phi_i(object[j]$, $s)$, $\ldots)$ of the work cycle (Figure~\ref{fig:contsubsys_example:case3}).

\Subsubsection{$\cs(thread, 2)$'s blocking times and request probabilities}
We complete the example in which we show how the contention subsystem represents the dependencies among the threads in the shared-object system.
We refer to an arbitrary job, say $job_i$, that $s=thread[n]$ carries out and explain how the contention subsystem's request probabilities $(\sR_{s})_{n\in [1,N]}$ and blocking periods $(\sB_{s})_{n\in [1,N]}$ represent the request probabilities, and respectively, the blocking periods in the shared-object system.
We justify this representation using the work cycle of $job_i$, when its path is in the path partition $\sP_j$, for every $j\in [1,3]$.
Note that for each such path partition, the period between two consecutive work cycles completed by $s$ is represented by $\sB_{s}(s)$ $=$ $\sT_{s}$ (Figure~\ref{fig:contsubsys_example}).
Namely, if $s$ carries out $job_i$ and consecutively $job_{i'}$, and their first demand requests are $\delta_i(s$, $d)$, and respectively, $\delta_{i'}(s$, $d')$, the blocking period $\sB_{s}(s)$ represents the period between these two events (in the contention subsystem).


%

\noindent{\bf The case of $(s, o_1, \bullet)$ paths.}
Consider the case where $s$ carries out $job_i$ with path $r$ $=$ $(s$, $o_1, \bullet)$, where $r$ $\in$ $\sP_2$, if $d$ is included in $r$ and $r$ $\in$ $\sP_3$, if $d$ is not included in $r$. 
We present the request probabilities among $s$, $relay(s$, $1)$ and $d$, as well as the blocking times on each of these items in the contention subsystem.

$\bullet$ {\em The probability $\sR_{s}(s$, $relay(s$, $1))$.}
The probability $\sR_{s}(s$, $relay(s$, $1))$ $=$ $R(s$, $o_1)$ (by the definition of $R$) denotes the contention subsystem event of $s$ demanding access to $relay(s$, $1)$,
which represents $s$ demanding access to $o_1$ immediately after $job_i$'s assignment in the shared-object system (figures~\ref{fig:contsubsys_example:case2} and \ref{fig:contsubsys_example:case3}, when $j = 1$).

$\bullet$ {\em The blocking period $\sB_{s}(relay(s$, $1))$.}
Let $W(s$, $o_1$, $d)$ denote the period during which $s$ blocks $o_1$, minus the possible blocking period on $d$ (after the supply of access to $o_1$) in the shared-object system.
Moreover, let $X(s$, $o_1$, $d)$ $=$ $\Pr[(s$, $o_1$, $d)]$ $\cdot$ $o_1.B[2]$ denote the (possible) blocking period of $s=thread[n]$ to $d=object[2]$, after gaining access to $o_1=object[1]$ in the shared-object system, 
where the probability $\Pr[(s$, $o_1$, $d)]$ $=$ $R(s$, $o_1)$ $\cdot$ $R(o_1$, $d)$ denotes the event of $s$ demanding access to $o_1$ and successively to $d$.
Namely, $X(s$, $o_1$, $d)$ is the time between the work cycle events $\delta_i(s$, $d)$ and $\phi_i(d$, $s)$ times the probability of $s$ to demand access to $d$ after gaining access to $o_1$ (Figure~\ref{fig:contsubsys_example:case2}).
Thus, in the shared-object system $W(s$, $o_1$, $d)$ $=$ $s.D[1]$ $-$ $X(s$, $o_1$, $d)$.
Therefore, $\sB_{s}(relay(s$, $1))$ $=$ $W(s$, $o_1$, $d)$ represents the period during which $s$ blocks $relay(s, 1)$ before possibly demanding access to $d$ in the contention subsystem.

$\bullet$ {\em The probabilities $\sR_{s}(relay(s$, $1)$, $d)$ and $\sR_{s}(relay(s$, $1)$, $s)$.}
Let $\sF_{(s,1,2)}$ and $\sF'_{(s,1,2)}$ denote the events in which $s$ demands access to $d$ after gaining access to $o_1$ in the shared-object system, and respectively, to $relay(s$, $1)$ in the contention subsystem.
The event $\sF'_{(s,1,2)}$ in the contention subsystem represents the event $\sF'_{(s,1,2)}$ in the shared-object system, and therefore, the probability of $\sF'_{(s,1,2)}$ is given by $\sR_{s}(relay(s$, $1)$, $d)$ $=$ $\Pr[(s$, $o_1$, $d)]$ (Figure~\ref{fig:contsubsys_example:case2}). 
Moreover, when the event $\sF'_{(s,1,2)}$ (and thus the event $\sF_{(s,1,2)}$) does not occur (Figure~\ref{fig:contsubsys_example:case3}), the respective job is completed and $s$ becomes idle or starts a new job with 
probability $\sR_{s}(relay(s$, $1)$, $s)$ $=$ $1$ $-$ $\sR_{s}(relay(s$, $1), d)$.

%

\noindent{\bf The case of $(s$, $d$, $\bullet)$ paths.}
Consider the case where $job_i$'s path is $(s, d, \bullet)\in \sP_1$.
In Figure~\ref{fig:contsubsys_example:case1},  $s$ demands access to $d$ immediately after it is assigned with $job_i$. 
This is represented in $\cs(thread$, $2)$ by $s$ demanding access to $d$, with probability $\sR_{s}(s$, $d)$ $=$ $R(s$, $d)$. 
Moreover, $s$ blocks $d$ for a period of $\sB_{s}(d)$ $=$ $s.B[2]$ in $\cs(thread$, $2)$, which represents the period between the events $\sigma_i(d$, $s)$ and $\phi_i(d$, $s)$ in figures~\ref{fig:contsubsys_example:case1} and \ref{fig:contsubsys_example:case2}. 
After the job completion and the release event of $d$ in $\cs(thread$, $2)$, $s$ enters, with probability $\sR_{s}(d$, $s)$ $=$ $1$, an idle period (of possibly zero length) until it starts carrying out a new job.

\noindent{\bf The case of $(s$, $o_3)$ paths.}
Consider the case where $s$ carries out $job_i$ with path $r = (s, o_3)\in \sP_3$.
In $\cs(thread$, $2)$, $s$ demands access to $relay(s$, $3)$, with probability $\sR_{s}(s$, $relay(s$, $3))$ $=$ $R(s$, $o_3)$,
which represents 
$s$ demanding access to $o_3$ immediately after $job_i$'s assignment in the shared-object system (Figure~\ref{fig:contsubsys_example:case3}).
The blocking period of $s$ on $relay(s$, $3)$ is $\sB_{s}(relay(s$, $3))$ $=$ $s.D[3]$, which in the shared-object system represents the time that $s$ is waiting to gain access to $o_3$ and then blocking it, i.e., the period between the work cycle events $\delta_i(s$, $o_3)$ and $\phi_i(o_3$, $s)$ (Figure~\ref{fig:contsubsys_example:case3}). 
After the job completion and the release event of $relay(s$, $3)$ in $\cs(thread$, $2)$, $s$ enters, with probability $\sR_{s}(relay(s$, $3)$, $s)$ $=$ $1$, an idle period (of possibly zero length) until it starts carrying out a new job. 

The contention subsystem $\cs(thread$, $d)$ $=$ $(\ch(thread$, $d)$, $(\sR_s)_{s\in thread}$, $(\sB_s)_{s\in thread})$, which we described above, represents the dependencies among the thread set, $thread$, and $d=object[2]$ in the shared-object system.


\Subsection{The case of systems with $M$ objects}
\label{s:appendix:proof}
In this section we prove that Theorem~\ref{thm:BaynatDallery} follows from lemmata~\ref{lem:cs_representation}, \ref{lem:framework_solution} and~\ref{lem:framework_complexity} (Corollary~\ref{thm:CorolBaynatDallery}). 

%
\begin{figure*}[t!]

\begin{tabular}{ccc}
\multicolumn{3}{ c }{
\begin{subfigure}[b]{\textwidth}
\centering
\includegraphics[width=0.35\textwidth]{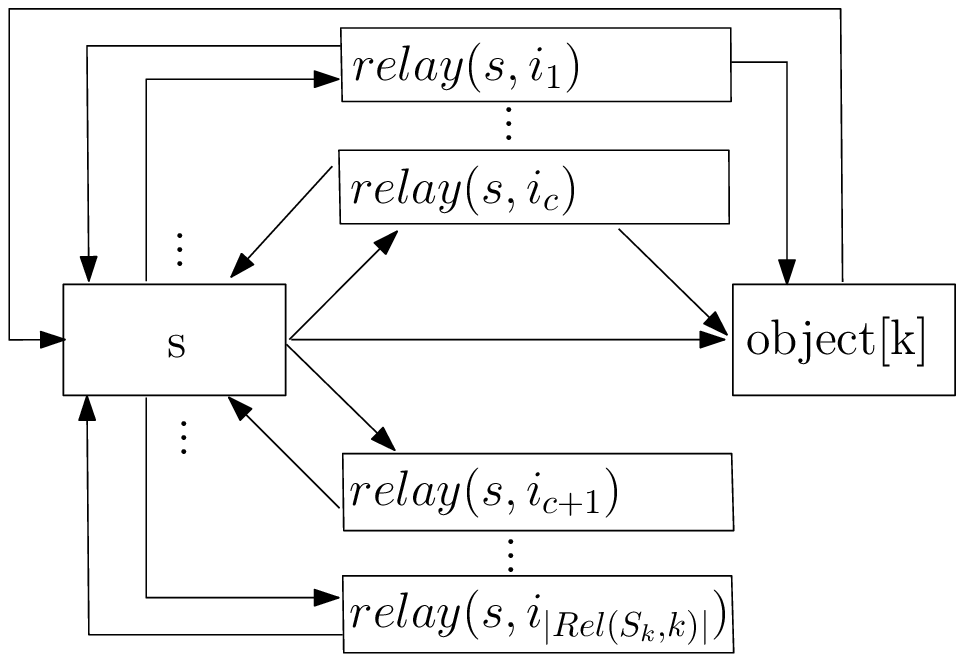}
\caption{The $\ch(S_k,k)$ contention graph, where $[i_1, i_c] = Rel(S_k,k) \setminus [k+1,M]$}
\label{fig:contgraph_general_case:graph}
\end{subfigure}
}
\\ 
\begin{subfigure}[b]{0.33\textwidth}
\centering
\includegraphics[width=0.9\textwidth]{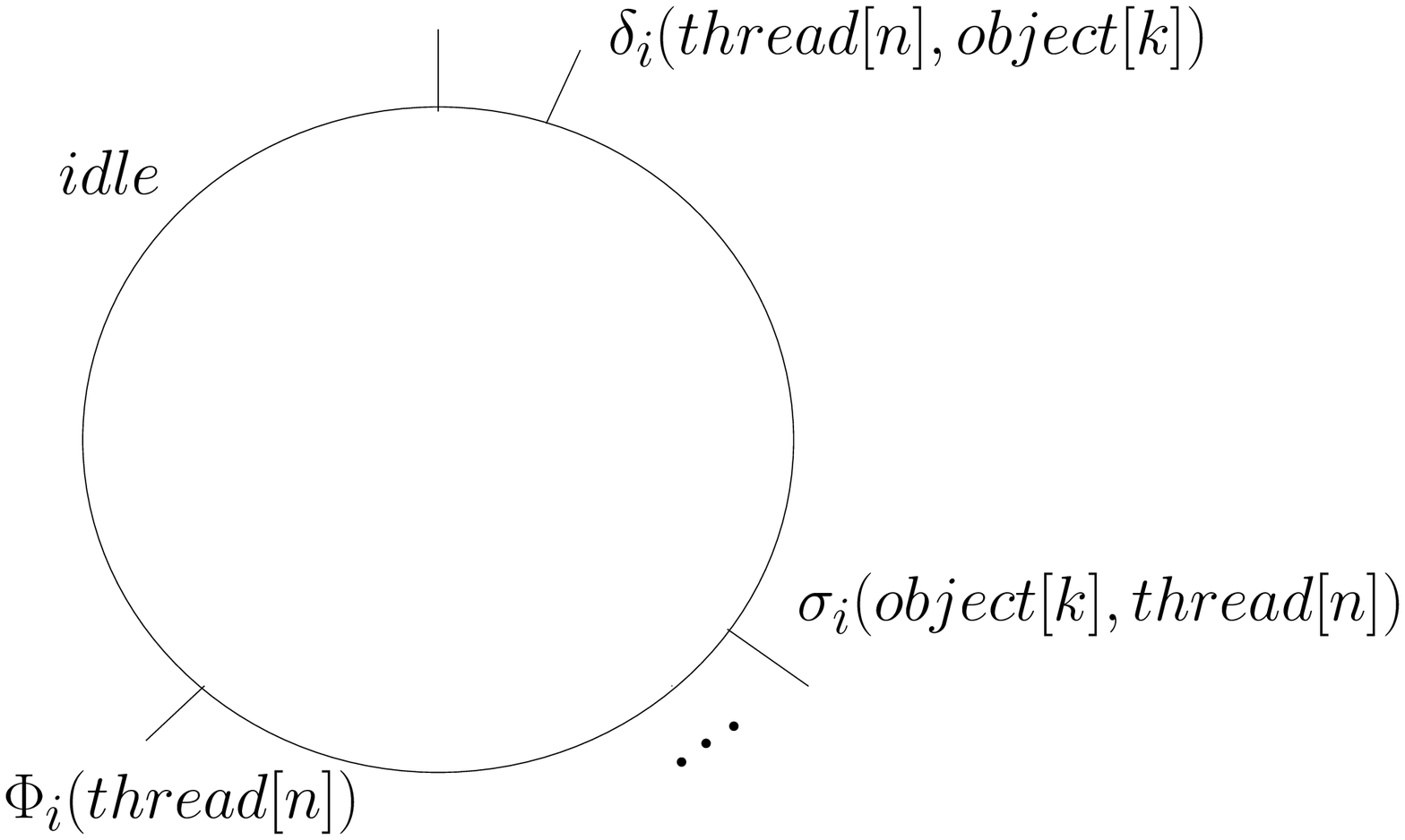}
\caption{Case 1 work cycles.}
\label{fig:contgraph_general_case:case1}
\end{subfigure} 
& 
\begin{subfigure}[b]{0.33\textwidth}
\centering
\includegraphics[width=0.9\textwidth]{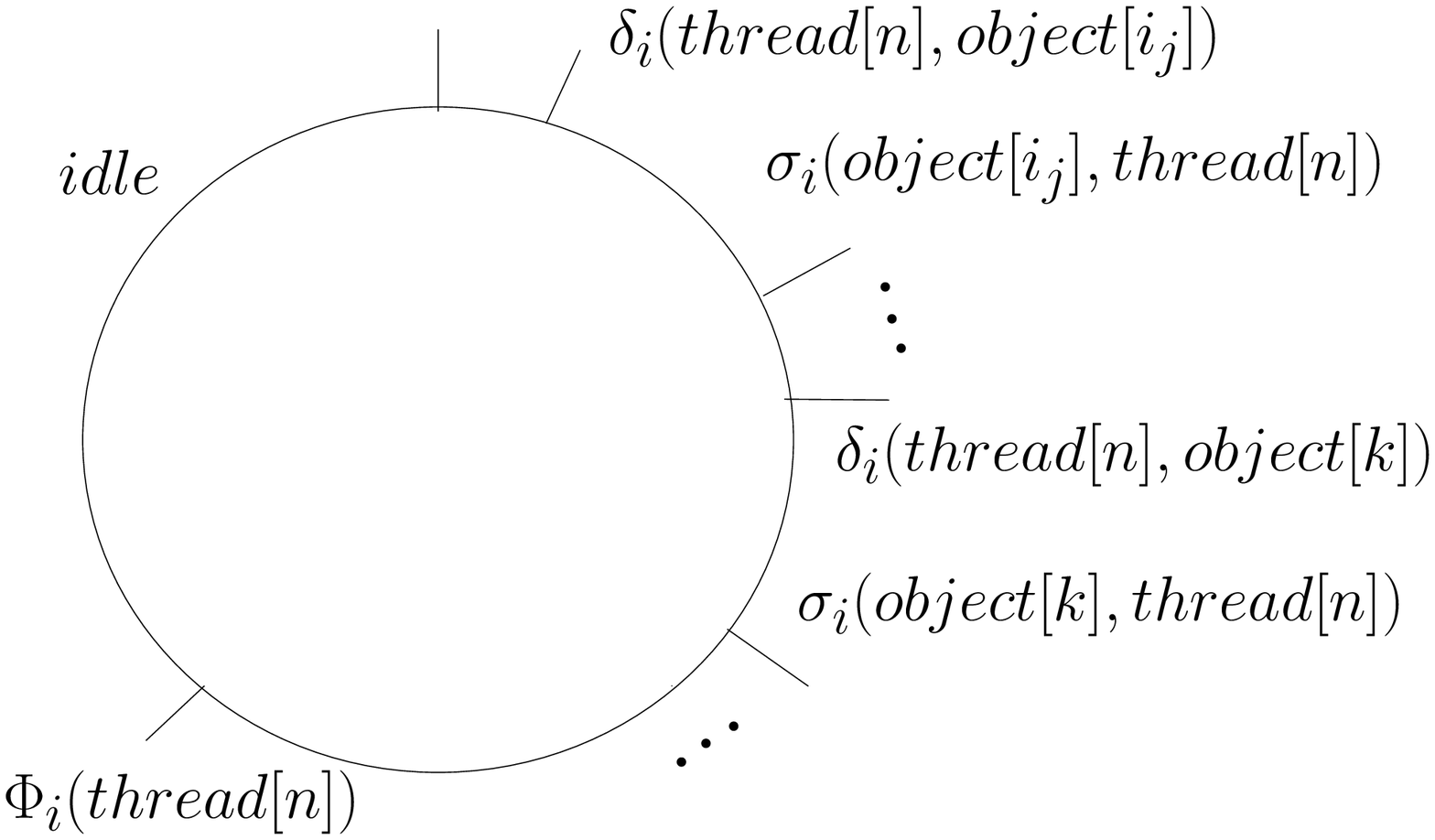}
\caption{Case 2 work cycles.}
\label{fig:contgraph_general_case:case2}
\end{subfigure} 
& 
\begin{subfigure}[b]{0.33\textwidth}
\centering
\includegraphics[width=0.9\textwidth]{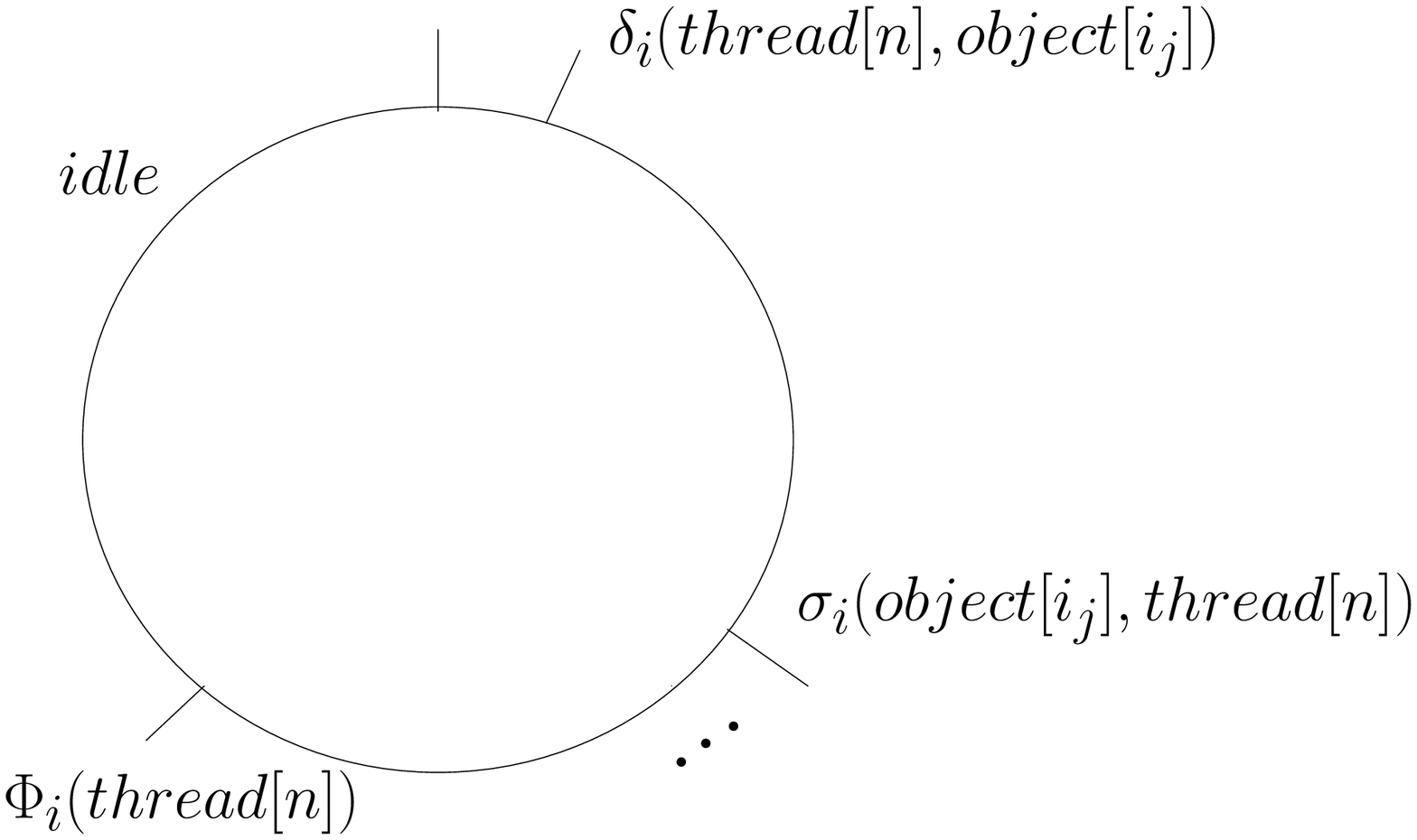}
\caption{Case 3 work cycles.}
\label{fig:contgraph_general_case:case3}
\end{subfigure}
\end{tabular}
\caption{The contention graph for $\cs(S_k, k)$ and the work cycles partitions,
$\sP(s, d)$ $=$ $\cup_{\ell\in [1,3]} \sP_\ell$, of $(\bullet$, $s$, $\bullet)$ paths,  where
$\sP_1 =$ $\{path$ $|$ $path$ $=$ $(\bullet$, $s$, $d$, $\bullet)\}$, 
$\sP_2$ $=$ $\{ path$ $|$ $path$ $=$ $(\bullet$, $s$, $object[i]$, $\bullet$, $d$, $\bullet)$ $\land$ $i\in Rel(S_k$, $k)$ $\setminus$ $[k+1,M]\}$, 
$\sP_3$ $=$ $\{path$ $|$ $path$ $=$ $(\bullet$, $s$, $\bullet)$ $\land$ $d$ $\notin$ $path\}$, $Rel(thread$, $k)$ $=$ $[1$, $M]$ $\setminus$ $\{k\}$ and $d = object[k]$. 
}
\label{fig:contgraph_general_case}
\end{figure*}

%

%

\begin{lemma}
\label{lem:cs_representation}
Consider a contention subsystem $\cs(S_k$, $k)$ $=$ $(\ch(S_k$, $k)$, $(\sR_s)_{s\in S_k}$, $(\sB_s)_{s\in S_k})$, where $S_k \in \{thread\}\cup s_k$ and $s_k = \{\{object[i]\}$ $|$ $i \in [1,k-1]\}$. Suppose that we are given the shared-object system's blocking and item inter-demand periods, as well as the request probabilities $R$.  It holds that $\cs(S_k$, $k)$ represents the dependencies among the threads in the shared-object system and the system's state. 
%

\end{lemma}

\proof{
%
We show a mapping of the shared-object system's state to the contention subsystem $\cs(S_k, k)$.
Given the shared-object system's blocking and item inter-demand periods, as well as the request probabilities, we construct the contention subsystem $\cs(S_k$, $k)$ $=$ $(\ch(S_k$, $k)$, $(\sR_s)_{s\in S_k}$, $(\sB_s)_{s\in S_k})$ based on the work cycles related to 
jobs with $(\bullet, s, \bullet)$ paths, where $s\in S_k$ and $s_k = \{\{object[i]\}$ $|$ $i \in [1,k-1]\}$.
We explain the representation of work cycles by the contention graph $\ch(S_k$, $k)$, as well as the representation of the shared-object system's request probabilities and state, i.e., blocking, pairwise inter-demand period and delay, by $(\sR_s)_{s\in S_k}$, and respectively, $(\sB_s)_{s\in S_k}$  in the contention subsystem.
This construction is the mapping that proves the lemma's statement.

The proof is organized as follows. 
In the first part, we construct the contention graph $\ch(S_k$, $k) = (\sV, \sE)$ using the work cycles related to $(\bullet, s, \bullet)$ paths. 
Moreover, in the second part we show that $(\sR_s)_{s\in S_k}$ and $(\sB_s)_{s\in S_k}$ represent the dependencies and the shared-object system's state with respect to 
$(\bullet, s, \bullet)$ paths.
In our construction, we assume the knowledge of the item inter-demand periods $\sT_v$, the system's state delay and blocking periods, as well as the request probabilities $R$.

\Subsubsection{\bf The graph $\ch(S_k, k)$ represents the work cycles related to the $(\bullet, s, \bullet)$ paths}
Let $\ch(S_k, k) = (\sV, \sE)$ be the contention graph of $\cs(S_k$, $k)$ (Figure~\ref{fig:contgraph_general_case:graph}) and consider $s$ to be an arbitrary element of $S_k$.
Recall that $Rel(thread$, $k)$ $=$ $[1$, $M]$ $\setminus$ $\{k\}$ and $Rel(\{object[\ell]\}$, $k)$ $=$ $[\ell+1,M]\setminus \{k\}$, where $\ell \in [1,k)$.
Moreover, let $\sP(s$, $object[k])$ $=$ $\cup_{\ell\in [1,3]} \sP_\ell$ be a path partition of $(\bullet$, $s$, $\bullet)$ paths, where
$\sP_1 =$ $\{path$ $|$ $path$ $=$ $(\bullet$, $s$, $object[k]$, $\bullet)\}$, 
$\sP_2$ $=$ $\{ path$ $|$ $path$ $=$ $(\bullet$, $s$, $object[\ell]$, $\bullet$, $object[k]$, $\bullet)$ $\land$ $\ell\in Rel(S_k$, $k)$ $\setminus$ $[k+1,M]\}$
and $\sP_3$ $=$ $\{path$ $|$ $path$ $=$ $(\bullet$, $s$, $\bullet)$ $\land$ $object[k]$ $\notin$ $path\}$.
We explain the representation of the thread work cycles for jobs with $(\bullet$, $s$, $\bullet)$ paths by the graph $\ch(S_k$, $k)$.

We define the elements of $\sV$ and $\sE$.
Let $\sV$ $=$ $\cup_{s\in S_k}\sV_{s}$, where $\sV_{s}$ $=$ $\{s$, $object[k]\}$ $\cup$ $\{relay(s,j)$ $|$ $j \in Rel(S_k,k)\}$.
The nodes $relay(s$, $j)$, $j\in Rel(S_k$, $k)$, are $s$'s distinct copies of a relay object, $object[j]$, which allow us to distinguish paths with respect to threads. 
The edges $\sE$ $=$ $\cup_{s \in S_k} \sE_{s}$ follow the three path partition cases, i.e., $\sE_s = \cup_{\ell\in [1,3]} \sE_\ell$,
where
$\sE_1$ $=$ $\{(s$, $object[k])$, $(object[k]$, $s)\}$,
$\sE_2$ $=$
$\{(s$, $relay(s$, $j)$, $(relay(s$, $j)$, $object[k])$, $(object[k]$, $s)\}$ and
$\sE_3$ $=$ $\{(s$, $relay(s$, $j))$, $(relay(s$, $j)$, $s)\}$.
Let $job_i$ be a job that $thread[n]$ carries out, such that item $s$ is either included in $job_i$'s object vector or $s = thread[n]$, and let $cycle(thread[n], job_i)$ be the respective work cycle. 
The edges in the sets $\sE_\ell$, where $\ell \in [1,3]$, represent the events in every possible $cycle(thread[n], job_i)$ after the supply of item $s$, if $s$ is an object, or after the assignment of $job_i$ to $s$, if $s = thread[n]$. 
For brevity, we refer to both events as the supply of item $s$.
The edge sets are defined according to the three sets of the path partition $\sP(s$, $object[k])$ $=$ $\cup_{\ell\in [1,3]} \sP_\ell$ as follows.
\begin{itemize}
\item $\sP_1$'s case refers to work cycles, for which $thread[n]$ demands access to $object[k]$, immediately after the supply of item $s$.
When $thread[n]$ gains access to $object[k]$, $job_i$ might require $thread[n]$ to demand access to another object, $object[k']$, where $k' \in [k+1, M]$. 
Upon $job_i$'s completion $thread[n]$ releases any acquired object (event $\Phi_i(thread[n])$ of the work cycle).
Thus, 
the edges $(s$, $object[k])$ and $(object[k]$, $s)$ of $\sE_1$ (Figure~\ref{fig:contgraph_general_case:graph})
represent the subvector
$(\delta_i(thread[n]$, $object[k]))$, and respectively, the subvector $(\sigma_i(object[k]$, $thread[n])$, $\ldots$, $\Phi_i(thread[n]))$
of the work cycle (Figure~\ref{fig:contgraph_general_case:case1}). 

\item $\sP_2$'s case refers to work cycles for which $thread[n]$, after the supply of item $s$, demands access to $object[j]$, where $j\in Rel(S_k,k)\setminus [k+1,M]$, and subsequently to $object[k]$.
Note that, by the definition of $\sP_2$, $thread[n]$ might also demand access to other objects in $(object[j]$, $object[k])$ $\cup$ $[object[k+1]$, $object[M$.
Thus, the edges $(s$, $relay(s$, $j)$, $(relay(s$, $j)$, $object[k])$, $(object[k]$, $s)$ of $\sE_2$ (Figure~\ref{fig:contgraph_general_case:graph})
represent the work cycle subvectors
$(\delta_i(thread[n]$, $object[j]))$, $(\sigma_i(object[j]$, $thread[n])$, $\ldots$ $\delta_i(thread[n]$, $object[k]))$, and respectively, $(\sigma_i(object[k]$, $thread[n])$, $\ldots$,\\ $\Phi_i(thread[n]))$ (Figure~\ref{fig:contgraph_general_case:case2}). 

\item $\sP_3$'s case refers to work cycles, for which $thread[n]$, after the supply of item $s$, demands access to $object[j]$, where $j\in Rel(S_k,k)$. 
Note that, by the definition of $\sP_3$, $thread[n]$ might also demand access to other objects in $(object[j]$, $object[M$, but not to $object[k]$.
Thus, the edges $(s$, $relay(s$, $j))$ and $(relay(s$, $j)$, $s)$ of $\sE_3$ (Figure~\ref{fig:contgraph_general_case:graph}),
represent the subvectors
$(\delta_i(thread[n]$, $object[j]))$, and respectively, $(\sigma_i(object[j])$, $thread[n])$, $\ldots$, $\Phi_i(thread[n]))$, where $j\in Rel(S_k,k)$ of the work cycle (Figure~\ref{fig:contgraph_general_case:case3}).
\end{itemize}


\Subsubsection{$(\sR_s)_{s\in S_k}$ and $(\sB_s)_{s\in S_k}$ represent the dependencies and the shared-object system's state with respect to $(\bullet,$ $s,$ $\bullet)$ paths}
%
We refer to an arbitrary job, say $job_i$, that $thread[n]$ carries out and explain how the contention subsystem's request probabilities $(\sR_{s})_{s\in S_k}$ and blocking periods $(\sB_{s})_{s\in S_k}$ represent the request probabilities, and respectively, the blocking periods in the shared-object system.
We verify this representation using the work cycle of $job_i$, when its path is in the path partition $\sP_j$, for every $j\in [1,3]$.
We remind that throughout this proof we refer to the supply of item $s$ as the supply of access to $object[\ell]$ for $thread[n]$, if $s=object[\ell]$, $\ell\in [1,k-1]$, or the assignment of $job_i$ to $thread[n]$, if $s = thread[n]$.

Note that for each such path partition, the period between two consecutive work cycles for jobs that include item $s$ is represented by $\sB_{s}(s)$ $=$ $\sT_{s}$ (Figure~\ref{fig:contsubsys_example}).
Namely, the blocking period $\sB_{s}(s)$ in the contention subsystem represents the period between the event of $thread[n]$ releasing item $s$ due to $job_i$, and consecutively, another thread, say $thread[n']$, releasing item $s$ due to a job, say $job_{i'}$, in the shared-object system.
We show this representation by looking into three cases of path partitions, i.e., 
(1) paths $(\bullet$, $s$, $object[k]$, $\bullet)$ in $\sP_1$,
(2) paths $r$ $=$ $(\bullet$, $s$, $object[j]$, $\bullet)$, where $j\in Rel(S_k,k)$ $\setminus$ $[k+1, M]$ 
($r$ $\in$ $\sP_2$, if $object[k]$ is included in $r$ and $r$ $\in$ $\sP_3$, if $object[k]$ is not included in $r$)
and
(3) paths $r$ $=$ $(thread[n]$, $\bullet$, $s$, $object[j]$, $\bullet)$, where $j\in Rel(S_k, k)$, and $object[k]$ $\notin$ $r$, i.e., $r\in \sP_3$.

\begin{enumerate}[(1)]

\item Consider the case where $job_i$'s path is in partition $\sP_1$,
i.e., $thread[n]$ carries out $job_i$ with path $(\bullet$, $s$, $object[k]$, $\bullet)$.
In Figure~\ref{fig:contgraph_general_case:case1},  $thread[n]$ demands access to $object[k]$ immediately after the supply of item $s$.
This event is represented in the contention subsystem by $thread[n]$ demanding access to $object[k]$, with probability $\sR_{s}(s$, $object[k])$ $=$ $R(s$, $object[k])$ $/$ $K(S_k$, $k)$ (by the definition of $R$), where $K(S_k$, $k)$ $=$ $\Sigma_{v\in Rel(S_k,k)\cup \{k\}}$ $R(s$, $object[v])$ is a normalizing constant .
Furthermore, $thread[n]$ blocks $object[k]$ for a period of $\sB_{s}(object[k])$ $=$ $s.B[k]$ in the contention subsystem, which represents the period between the events $\sigma_i(object[k]$, $thread[n])$ and $\Phi_i(thread[n])$ in the work cycles presented in figures~\ref{fig:contgraph_general_case:case1} and \ref{fig:contgraph_general_case:case2}. 
This is due to the fact that $s.B[k]$ also considers requests for accessing $object[k]$ while following either job path $(thread[n]$, $\bullet$, $s$, $object[k]$, $\bullet)$ or $(thread[n]$, $s$, $\bullet$, $object[k]$, $\bullet)$ (Lemma~\ref{lemma:blocking}). 
After the job completion and the release event of $object[k]$, another supply event of item $s$ occurs or it becomes idle.
Namely, if $s=thread[n]$, $s$ will either start carrying out the next pending job or it will become idle, and if $s=object[\ell]$, the thread on $s$'s queue top will gain access to $s$ in case the queue is not empty, otherwise $s$ will become idle. 
These are certain events in the shared-object system and therefore occur with probability $\sR_{s}(object[k]$, $s)$ $=$ $1$ in the contention subsystem. 

\item Consider the case where $thread[n]$ carries out $job_i$ with path $r$ $=$ $(\bullet$, $s$, $object[j]$, $\bullet)$, where $j\in Rel(S_k,k)$ $\setminus$ $[k+1, M]$.
Note that $r$ $\in$ $\sP_2$, if $object[k]$ is included in $r$ and $r$ $\in$ $\sP_3$, if $object[k]$ is not included in $r$. 
We present the request probabilities among $s$, $relay(s$, $j)$ and $object[k]$, $j\in Rel(S_k, k)$ $\setminus$ $[k+1,M]$, as well as the blocking times on each of these items in the contention subsystem.
We present (i) the probability $\sR_{s}(s$, $relay(s$, $j))$, (ii) the blocking period $\sB_{s}(relay(s$, $j))$ and (iii) the probabilities $\sR_{s}(relay(s$, $j)$, $object[k])$ and $\sR_{s}(relay(s$, $j)$, $s)$.
\begin{enumerate}[(i)]
\item The probability $\sR_{s}(s$, $relay(s$, $j))$ $=$ $R(s$, $object[j])$ $/$ $K(S_k$, $k)$ (by the definition of $R$) denotes the contention subsystem event of $thread[n]$ demanding access to $relay(s$, $j)$ after the supply of item $s$,
which represents $thread[n]$ demanding access to $object[j]$ immediately after the supply of item $s$ in the shared-object system (figures~\ref{fig:contgraph_general_case:case2} and \ref{fig:contgraph_general_case:case3}, when $j$ $\in$ $Rel(S_k$, $k)$ $\setminus$ $[k+1$, $M]$).

\item 
Consider the case where $thread[n]$ might demand access to objects in $(object[j]$, $object[M$, 
after the supply of $object[j]$ in the shared-object system.
Let $W(s$, $object[j]$, $object[k])$ as the period during which $thread[n]$ blocks $object[j]$ minus $thread[n]$'s possible blocking period on $object[k]$ in the shared-object system, due to a path in $\sP_2$. 
We refer to
$X(s$, $object[j]$, $object[k])$ $=$ $\Sigma_{\ell=j}^{k-1}$ $\Pr[(s$, $object[j]$, $\bullet$, $object[\ell]$, $object[k])]$ $\cdot$ $object[\ell].B[k]$ 
as the possible blocking period of $thread[n]$ to $object[k]$, after gaining access to $object[j]$ in the shared-object system. 
Note that the probability 
$\Pr[(s$, $object[j]$, $\bullet$, $object[\ell]$, $object[k])]$ $=$ $R(s$, $object[j])$ $\cdot$ $[\Sigma_{k=1}^{j-\ell}$ $R^k(object[\ell]$, $object[j])]$ $\cdot$ $R(object[\ell]$, $object[k])$
denotes the event in which $thread[n]$, after the supply of item $s$, demands access to $object[j]$, possibly to other objects in $(object[j]$, $object[\ell])$, to $object[\ell]$ and successively to $object[k]$. 
Namely, let $X(s$, $object[j]$, $object[k])$ denote the time between the work cycle events $\delta_i(thread[n]$, $object[k])$ and $\Phi_i(thread[n])$ times the probability of $thread[n]$ to demand access to $object[j]$, after gaining access to $s$, and subsequently to $object[k]$ (Figure~\ref{fig:contgraph_general_case:case2}).
Thus, in the shared-object system $W(s$, $object[j]$, $object[k])$ $=$ $s.D[j]$ $-$ $X(s$, $object[j]$, $object[k])$.
Therefore, the period during which $thread[n]$ blocks $relay(s, j)$, after the supply of item $s$, and before possibly demanding access to $object[k]$ in the contention subsystem is represented by
$\sB_{s}(relay(s$, $j))$ $=$ $W(s$, $object[j]$, $object[k])$.

\item Let $\sF_{(s,j,k)}$ and $\sF'_{(s,j,k)}$ denote the event in which $thread[n]$, demands access to $object[j]$, after the supply of item $s$, and subsequently to $object[k]$ in the shared-object system, and respectively, the event in which $thread[n]$, after the supply of item $s$, demands access to $relay(s$, $j)$ and consecutively to $object[k]$ in the contention subsystem.
Note that $thread[n]$ might demand access to other objects in $(object[j]$, $object[k])$ in the shared-object system.
The event contention subsystem $\sF'_{(s,j,k)}$ represents the event $\sF'_{(s,j,k)}$ in the shared-object system, and therefore, the probability of $\sF'_{(s,j,k)}$ is given by $\sR_{s}(relay(s$, $j)$, $object[k])$ which equals to $\Pr[(s$, $object[j]$, $object[k])]$ (Figure~\ref{fig:contgraph_general_case:case2}). 
Moreover, when the event $\sF'_{(s,j,k)}$ (and thus the event $\sF_{(s,j,k)}$) does not occur (Figure~\ref{fig:contgraph_general_case:case3}), the respective job is completed and $thread[n]$ becomes idle or starts a new job with 
probability $\sR_{s}(relay(s$, $j)$, $s)$ $=$ $1$ $-$ $\sR_{s}(relay(s$, $j)$, $object[k])$.
\end{enumerate}

\item Consider the case where $thread[n]$ carries out $job_i$ with path $r$ $=$ $(thread[n]$, $\bullet$, $s$, $object[j]$, $\bullet)$, where $j\in Rel(S_k, k)$, and $object[k]$ $\notin$ $r$, i.e., $r\in \sP_3$.
In the contention subsystem, $thread[n]$, after the supply of item $s$, demands access to $relay(s$, $j)$, with probability $\sR_{s}(s$, $relay(s$, $j))$ $=$ $R(s$, $object[j])$ (by the definition of $R$),
which represents 
$thread[n]$ demanding access to $object[j]$ immediately after the supply of item $s$ in the shared-object system (Figure~\ref{fig:contgraph_general_case:case3}).
The blocking period of $thread[n]$ on $relay(s$, $j)$ is $\sB_{s}(relay(s$, $j))$ $=$ $s.D[j]$, which in the shared-object system represents the time that $thread[n]$ is waiting to gain access to $object[j]$, blocks it while it that does read well here. Too many 'it'. Please use the name $thread[n]$ and $object[j]$. Another way to go is to use (conditional) events... the time between $\delta_{i,n}(d|s)$ and $\ldots$.  possibly demands access to other objects (except for $object[k]$) and releases all of its acquired objects.
Namely, $\sB_{s}(relay(s$, $j))$ represents the period between the work cycle events $\delta_i(thread[n]$, $object[j])$ and $\Phi_i(thread[n])$, after the supply of item $s$ (Figure~\ref{fig:contgraph_general_case:case3}). 
After the job completion and the release event of $relay(s$, $j)$ in the contention subsystem, either another thread gains access to item $s$ or item $s$ becomes idle (no thread is accessing it), i.e., this event is certain to happen in the shared-object system.

\end{enumerate}

The $\cs(S_k$, $k)$ $=$ $(\ch(S_k$, $k)$, $(\sR_s)_{s\in S_k}$, $(\sB_s)_{s\in S_k})$ contention subsystem, which we described above, represents the dependencies among the item set, $S_k$, and $object[k]$ in the shared-object system. 
Therefore, the proof is complete.
}
\begin{lemma}
\label{lem:framework_solution}
The framework of Baynat and Dallery can approximate the delay $s.D[k]$ and inter-demand period $s.T[k]$ through the contention subsystem $\cs(S_k, k)$.
\end{lemma}

%
\proof{
We first give the definition of a \emph{Ramesh-Perros subsystem} (RPS) that is introduced in~\cite{DBLP:journals/pe/RameshP01} and show that a contention subsystem can be directly mapped to an RPS. 
Ramesh and Perros show that we can find the pairwise inter-demand period, and delay of (blocking) communications in an RPS using a framework proposed by Baynat and Dallery in~\cite{DBLP:journals/pe/BaynatD96}.
Thus, 
we can use the Baynat-Dallery framework to estimate the values of $s.T[k]$ and $s.D[k]$, for every $s\in S_k$, given a contention subsystem $\cs(S_k,k)$ as an input. 
In Appendix~$\ref{s:appendix:framework_details}$, we present an adapted version of the Baynat-Dallery framework (Algorithm~\ref{alg:framework}) and explain the calculation of the delay $s.D[k]$ and inter-demand period $s.T[k]$, given the contention subsystem $\cs(S_k, k)$.

A Ramesh-Perros subsystem is defined as follows.
%
%
%
Let $\{o[1]$, $\ldots$, $o[M]\}$ be a set of servers, such that tier-$k$ includes only server $o[k]$, where $k \in [1,M]$, and $\{t[1]$, $\ldots$, $t[N]\}$ be a set of clients.
Moreover, let
$s_k = \{ \{o[1]\}, \ldots, \{o[k-1]\}\}$ (i.e., $s_k=\emptyset$ when $k=1$) and
$S_k$ $\in$ $\{\{t[1]$, $\ldots$, $t[N] \}\}$ $\cup$ $s_k$.
A Ramesh-Perros subsystem, is denoted by $\rp(\mathtt{S}_k$, $k)$ $=$ $(\mathtt{H}(S_k$, $k)$, $(\mathtt{R}_x)_{x\in S_k}$, $(\mathtt{B}_x)_{x\in S_k})$ and defined as follows:

\begin{enumerate}[(1)]
\item  The {\em RPS graph}
$\mathtt{H}(S_k$, $k)= (\cV, \cE)$ has the set of vertices $\cV$ $=$ $\cup_{x\in S_k}\cV_{x}$,
and the set of edges $\cE$ $=$ $\cup_{x\in S_k}$ $\cE_{x}$, such that $\cV_{x}$ $=$ $\{x\}$ $\cup$ $\{o[k]\}$ $\cup$ $\{relay(x,j)$ $|$ $j$ $\in$ $\mathtt{Rel}(S_k$, $k)\}$ and $\cE_{x}$ $=$ $\cE^1_{x}$ $\cup$ $\cE^2_{x}$ $\cup$ $\cE^3_{x}$,
where
$\mathtt{Rel}(\{t[1]$, $\ldots$, $t[N]\}$, $k)$ $=$ $[1,M]\setminus \{k\}$,
$\mathtt{Rel}(\{o[i]\}$, $k)$ $=$ $[i+1,M]\setminus \{k\}$, where $o[i] \in s_k$,
$\cE^1_{x}$ $=$ $\{(x$, $o[k])$, $(o[k]$, $x)\}$,
$\cE^2_{x}$ $=$  $\cup_{j\in \mathtt{Rel}(S_k,k)\setminus [k+1,M]}\{(x$, $relay(x$, $j))$, $(relay(x,j)$, $o[k])$, $(o[k]$, $x)\}$, and
$\cE^3_{x}$ $=$  $\cup_{j\in \mathtt{Rel}(S_k,k)}\{(x$, $relay(x$, $j))$, $(relay(x$, $j)$, $x)\}$.
Note that $\mathtt{H}(S_k$, $k)$ is a simple directed graph, i.e., there are no multiple edges between two vertices.

\item The {\em RPS request probability matrices} $\mathtt{R}_x$ for $\mathtt{H}(S_k$, $k)$ $=$ $(\cV$, $\cE)$ and $x \in S_k$, where, $\mathtt{R}_x[v_k,v_\ell]$ is the probability that the process at vertex $v_u \in \mathtt{V}_x$ forwards the client request to the server at vertex $v_\ell \in \mathtt{V}_x$, for an edge $(v_u$, $v_\ell)$ in $\cE_x$.

\item The {\em RPS blocking periods}, $(\mathtt{B}_x)_{x\in S_k}$, where $\mathtt{B}_x$ is a function over $\mathtt{V}_x$, and $x\in S_k$ refers to client or server processes.
Note that the edges in $\cE_x^i$ form a directed circle in the graph $\mathtt{H}(S_k,k)$, where $x\in S_k$ and $i\in [1,3]$.
A client request of $x \in \mathtt{V}_x$ to server $o[k]$, is blocking $o[k]$ for $\mathtt{B}_x(o[k])$ $=$ $x.\mathtt{B}[k]$ time.
A client request of $x \in \mathtt{V}_x$ to server $relay(x,j)$, blocks that server for a period of $\mathtt{B}_x(relay(x$, $j))$ $=$ $x.\mathtt{D}[j]$, if $j\in [k+1,M]$ and $\mathtt{B}_x(relay(x$, $j))$ $=$ $\mathtt{W}(x$, $relay(x,j)$, $o[k])$ otherwise ($j\in \mathtt{Rel}(S_k,k)\setminus [k+1,M]$).
Once the servers return to process $x \in \mathtt{V}_x$ with an answer, after a period of $\mathtt{B}_x(x)$ $=$ $\mathtt{T}_x$, process $x$ sends a new client request to a server in $\mathtt{V}_x \setminus \{x\}$.
Note that $x.\mathtt{B}[k]$, $x.\mathtt{D}[j]$, $\mathtt{W}(x$, $relay(x,j)$, $o[k])$ and $\mathtt{T}_x$ are functions of $x$ and $k$, $x$ and $j$, $x$, $relay(x,j)$ and $o[k]$, and respectively, $x$.
\end{enumerate}

Note that this definition of an RPS is a special case of the definition given in~\cite{DBLP:journals/pe/RameshP01}, which is adapted to our purposes.
A contention subsystem $\cs(S_k,k)$ is directly mapped to an RPS $\rp(\mathtt{S}_k,k)$, by setting
$o[k] = object[k]$, 
$t[n] = thread[n]$,
a message to be a demand request,
$\mathtt{H}(S_k$, $k)$ $=$ $\ch(S_k, k)$, 
$(\mathtt{R}_x)_{x\in S_k}$ $=$ $(\sR_s)_{s\in S_k}$ and
$(\mathtt{B}_x)_{x\in S_k}$ $=$ $(\sB_s)_{s\in S_k}$,
where $k\in [1,M]$ and $n\in [1,N]$.
Namely, we set 
$x.\mathtt{B}[k]$ $=$ $s.B[k]$, 
$x.\mathtt{D}[j]$ $=$ $s.D[j]$, 
$\mathtt{W}(x$, $relay(x,j)$, $o[k])$ $=$ $W(x$, $relay(s,j)$, $object[k])$ and 
$\mathtt{T}_x$ $=$ $\sT_{s}$,
where $x \in \mathtt{S}_k$ equals to the respective $s \in S_k$ following the mapping that we described above.
%
%
}

\begin{lemma}
\label{lem:framework_complexity}
The running time of each framework iteration is $O(M \cdot N^4)$.
\end{lemma}

\proof{
The running time per iteration of the Baynat and Dallery algorithm~\cite{DBLP:journals/pe/BaynatD96} is in $O(|stations|\cdot$ $|classes|^3)$, where $|stations|$ and $|classes|$ are the numbers of the network's stations and classes, respectively. 
In the context of shared-object systems, the number of $stations$, $|stations|$, corresponds to the number of vertices $|\mathcal{V}|$ of the contention graph $\ch(S_k, k)$ $=$ $(\mathcal{V}$, $\mathcal{E})$ and the number of classes to $S_k$'s cardinality, where $k\in [1,M]$. 
Thus, the running time of each iteration is in $O(|\mathcal{V}| \cdot |S_k|^3)$. 
Notice that $|\mathcal{V}| = N + N\cdot (M-1) + 1$ and $|S_k| = N$, if $S_k = thread$, and $|\mathcal{V}| = M - \ell + 1$ and $|S_k| = 1$, if $S_k = \{object[\ell]\}$, where $\ell\in [1,k-1]$. 
The result follows by taking the maximum $|\mathcal{V}|$ and $|S_k|$ of these two cases.
}

\begin{corollary}
\label{thm:CorolBaynatDallery}
Lemmata~\ref{lem:cs_representation}, \ref{lem:framework_solution} and~\ref{lem:framework_complexity} imply Theorem~\ref{thm:BaynatDallery}.
\end{corollary}


\Section{Finding an $\varepsilon$-OSE}
%
\label{s:appendix:ose}
We present a procedure (Algorithm~\ref{alg:yOSEf}) for finding $\varepsilon$-OSEs.
We give a detailed explanation of Algorithm~\ref{alg:yOSEf} (Section~\ref{s:appendix:alg_descr}) and analyze its running time (Lemma~\ref{th:running_time_yOSEf} of Section~\ref{s:appendix:alg_compl}).
We also detail the algorithm's functions (sections~\ref{s:appendix:alg_init}, \ref{s:appendix:alg_upd}, and  respectively,~\ref{s:appendix:alg_recalcs}), which are $initializeSystemState()$, $updateStates()$, and $recalcB()$ and $recalc\sT()$.

\Subsection{The $\varepsilon$-OSE solver}
\label{s:appendix:alg_descr}
%
%
The procedure always halts and computes an approximated equilibrium, $\varepsilon$-OSE, when such is reachable.  
Namely, whenever the job arrival and completion rates become equal, the procedure returns the system state in an $\varepsilon$-OSE, or indicates that an OSE is not a state that the system can be in.
The procedure sets initial values to the system state, $c[s,d]$, and then uses the proposed methods (sections~\ref{s:RoutingBlocking} to \ref{s:cont_graphs}) for estimating $c[s,d]$ iteratively, until convergence.
The decision on when to stop considers the system inter-demand period, $\{\sT_{item}\}_{item\in V \setminus \{object[M]\}}$, and stops whenever there is no $item\in V\setminus \{object[M]\}$ for which the change in $\sT_{item}$ is greater than $\varepsilon$ since the previous iteration, where $\dgraph$ $=$ $(V,E)$ in the acquisition graph.

The procedure's input includes the system parameters, i.e., 
%
%
number of threads $N$ and objects $M$, the jobs, $job_i$, $i\in [1,J]$, and their arrival rates $\{\lambda_{i,n}\}_{i\in [1,J], n\in [1,N]}$ to each $thread[n]$, $n \in [1,N]$, as well as the request probability matrix, $R$.
The procedure's output includes the delay $D$, inter-demand period $T$ and blocking period $B$ between all system items, as well as, the item inter-demand periods $\sT_v$, for every item $v\neq object[M]$.

The procedure starts with a system state that represents the case in which all queues are empty (see the function $initializeSystemState()$). It then estimates the state of a system in which threads can block one another, and the delay grows as more requests are pending in the queues.
The main part of the pseudocode (Algorithm~\ref{alg:yOSEf}) consists of a repeat-until loop (lines \ref{alg:yOSEf:line:repeat}--\ref{alg:yOSEf:line:until}) that follows the procedure's initialization (line \ref{alg:yOSEf:line:init}).
Before the procedure can return its output value, the loop has to end either when the $ \varepsilon$-OSE conditions are satisfied or when the procedure detects that an OSE cannot be reached. 
%
%
%
Each iteration aims at further improving the $\varepsilon$-OSE estimation.
The repeat-until loop then exits when no item changes by at least $\varepsilon$ between every two iterations (and thus the system state satisfies the conditions of an approximated equilibria).

In every iteration, the procedure computes (i) the blocking periods of demand requests for objects, $s.B[k]$ (the function $updateStates()$), (ii) the thread inter-demand periods, $\sT_{thread[n]}$ (the function $augmntThreadBlock()$), and (iii) the object inter-demands, $\sT_{object[j]}$ (which is the function $updateStates()$), where $(s$, $object[k])$ is an edge of the acquisition graph $\dgraph$, $n\in [1,N]$ and $j\in [1,M-1]$.
The repeat-until loop repeats the steps (i), (ii) and (iii), which deals with interdependencies using alternating backward and forward iterations. Namely, it resolves the forward dependencies in which $(s$, $object[k])$'s blocking period, $s.B[k]$, depends on $object[\ell]$'s delay by iterating backward, where $\ell \in (k,M]$, i.e., starting from $k=M$ and counting downwards, we can estimate $s.B[k]$, because (in a system that its state satisfies the equilibrium conditions) all of $(s,object[k])$'s forward dependencies can be resolved. Similarly, its uses forward iterations for resolving backward dependencies with respect to $d$'s item inter-demand period, $\sT_{object[k]}$, because all of $object[k]$'s backward dependencies are resolved.
Moreover, the function $augmntThreadBlock()$ allow the repeat-until loop to stop whenever the job inter-arrival time becomes less or equal than the time it takes that thread to complete such jobs, i.e., there's no OSE.

The blocking period estimation, step (i), starts from the last system object, $object[M]$, where there are no dependencies on the delay of subsequent demand requests.
For every $s$ such that $(s,object[M])$ is an edge of $\dgraph$, we calculate the delay, $s.D[M]$, and the pairwise inter-demand period, $s.T[M]$, of demand requests to $object[M]$ through the $updateStates(\#B$, $k$, $thread$, $object$, $\{\sT_v\}_{v\in allStates}$, $R)$ function (line \ref{alg:yOSEf:line:blocking}), where $allStates$ $=$ $\{thread[n]$ $|$ $n\in [1,N]\}$ $\cup$ $\{object[k]$ $|$ $k\in [1,M-1]\}$.
Therefore, the procedure can compute the blocking period of demand requests to $object[M-1]$, $s.B[M-1]$, because it has just estimated the dependencies of subsequent demand requests for $object[M]$.
Repeating this process for $k = M-1, \ldots, 1$, the procedure compute the blocking periods of demand requests to any $object[k]$, $s.B[k]$, where $k\in [1,M]$.
Note that this is possible, since in every step $k$, $k = M-1, \ldots, 1$, of this for loop, we have already computed the demand request delays and the inter-demand periods of every $object[k']$, $object[k].D[k']$, and respectively, $object[k].T[k']$, where $k' \in [k+1,M]$ (Section~\ref{s:blocking}).
%

%
After computing the blocking periods, delays and pairwise inter-demand periods, the procedure estimates the threads' inter-demand periods (Section~\ref{s:appendix:requestGenerationTime}), which can be used to estimate the job completion rates.
It does this in step (ii), through the function $augmntThreadBlock(\Sigma_i\lambda_{i,n}$, $blocking(n))$ (line \ref{alg:yOSEf:line:thread_updates}) and by using the thread idle probabilities (Appendix~\ref{s:appendix:jobAssignmentProbability}), where $blocking(n)$ $=$ $A$ $+$ $\Sigma_{k = 1}^M R(thread[n]$, $object[k])$ $\cdot$ $thread[n].D[k]$.
%
%
Moreover, note that the $augmntThreadBlock()$ function (Section~\ref{ss:tidp}) allows the repeat-until loop to stop whenever it detects that the inter-arrival time of jobs to a thread becomes less or equal than the time it takes that thread to complete such jobs (i.e., there's no OSE). Note that the repeat-until loop breaks when the Boolean variable $loopEnd$ is true (line~\ref{alg:yOSEf:line:thread_updates}).

%
Step (iii) uses the $updateStates()$ function (line~\ref{alg:yOSEf:line:throughput}) for calculating the object inter-demand periods, after calculating the new estimates for the delay and the pairwise inter-demand periods.
The procedure estimates the inter-demand period $\sT_{object[k]}$, for each $object[k]$, $k \in [1,M-1]$, via the item inter-demand period of demand requests for items that precede $object[k]$ in a job path $(\bullet$, $item[j]$, $object[k]$, $\bullet)$ (Section~\ref{ss:object_inter_demand_periods}).
%
%
Therefore, the procedure estimates the inter-demand periods for each $object[k]$ in the order of $k = 1, \ldots$, $M-1$.
%
%
%

Once the procedure verifies the satisfaction of the $\varepsilon$-OSE conditions (Section~\ref{s:preliminaries}), the repeat-until loop end 
(line~\ref{alg:yOSEf:line:convergence}) and the procedure returns.

\Subsection{The $initializeSystemState()$ function}
\label{s:appendix:alg_init}
The procedure $initializeSystemState()$ initializes Algorithm~\ref{alg:yOSEf} assuming that there is no contention, i.e., all queues have zero length in the shared-object system (lines~\ref{alg:yOSEf:line:init} and~\ref{alg:yOSEf:line:init_start}).
%
%
In the context of shared-object systems, no contention means that the delay of each demand request equals its blocking period, $s.D[k] = s.B[k]$, for all items $s$ and $d=object[k]$.
%
%
Thus, the procedure initializes the blocking periods $s.B[k]$ through the function $initRecord()$ (lines~\ref{alg:yOSEf:line:init_s_start}--\ref{alg:yOSEf:line:init_s_end}). It uses the average demand completion period $f_{s,object[k]}$, as proposed in Section~\ref{ss:job_compl_per}, which is a constant.
That is, $s.B[M]$ $=$ $A$ $+$ $R(s,d)$ $\cdot$ $R(d,d)$ $\cdot$ $f_{s,d}$, if $d = object[M]$ and $s.B[k]$ $=$ $A$ $+$ $R(s,d)$ $\cdot$ $R(d,d)$ $\cdot$ $f_{s,d} + \Sigma_{\ell=k+1}^M R(d, object[\ell]) \cdot  d.D[\ell]$, if $d = object[k]$, $k\in [1,M-1]$, where $s$ $\in$ $\{thread[n]$ $|$ $n\in [1,N]\}$ $\cup$ $\{object[i]~|$ $i \in [1,k-1]\}$.
Namely, the blocking period of a request to access (demand) the last object equals to the acquisition period plus the average demand completion period on that object and the blocking period of a demand to $d=object[k]$, $k\in [1,M-1]$, is recursively computed as the sum of the acquisition period, $A$, the average demand completion period $R(s,d)$ $\cdot$ $R(d,d)$ $\cdot$ $f_{s,d}$ plus the average blocking period of a subsequent demand to one of the following objects $d.B[\ell]$, $\ell \in [k+1,M]$, weighted by the probability of sending such a demand, $R(d, object[\ell])$, $\ell \in [k+1,M]$.
%
Moreover, 
the $s.T[k]$ pairwise inter-demand periods are set to equal $s.B[k]$.

The $\sT_{thread[n]}$ (thread) inter-demand periods are computed (line~\ref{alg:yOSEf:line:init_T_thread}) through the function $augmentThreadBlock(\Sigma_i \lambda_{i,n}$, $blocking(n))$, where $blocking(n)$ is defined in line~\ref{alg:yOSEf:line:thread_T}.
Furthermore, the $\sT_{object[j]}$ (object) inter-demand periods are set to $\Sigma_{\ell=j+1}^M R(object[j]$, $object[\ell])$ $\cdot$ $object[j].T[\ell]$ (line~\ref{alg:yOSEf:line:init_T_object}).
Algorithm~\ref{alg:yOSEf} iteratively finds the correct values of the pairwise and item inter-demand periods (arbitrary initialization of the system's state is proposed in~\cite{DBLP:journals/pe/RameshP01}).

\Subsection{The $updateStates()$ function}
\label{s:appendix:alg_upd}
This function updates the blocking periods and the item inter-demand periods with respect to $object[k]$ (lines \ref{alg:yOSEf:line:proc_start}--\ref{alg:yOSEf:line:proc_end}).
If the input tag is $\#B$, the function updates the blocking periods $s.B[k]$ for every $s$ $\in$ $\{thread[n]$ $|$ $n\in [1,N]\}$ $\cup$ $\{object[j]$ $:$ $j \in [1,k-1]\}$, as in Section~\ref{s:blocking}.
Otherwise, if the input tag is $\#\sT$, the function updates the inter-demand period of $d=object[k]$, $\sT_d$, as in Section~\ref{s:throughput}.
%

%
%

The procedure $updateStates()$ defines the contention subsystem for every set of item sources $S_k$ using the function $defContentionSubsystem()$ (line~\ref{alg:yOSEf:upd_st_def_thr} if $S_k = thread$ and line~\ref{alg:yOSEf:line:def_obj_cs} if $S_k = \{object[j]\}$, $j\in [1,k-1]$).
It then calculates the delay, $s.D[k]$, and pairwise inter-demand period, $s.T[k]$, for every $s\in S_k$ using the $BDF()$ function of Appendix~\ref{s:appendix:framework_details} (line~\ref{alg:yOSEf:line:estCont_thr} if $S_k = thread$ and line~\ref{alg:yOSEf:line:estCont_obj} if $S_k = \{object[j]\}$, $j\in [1,k-1]$).
The procedure ends with the computation of the blocking periods or the inter-demand periods, depending on the $tag$ with which the procedure was called.


\Subsection{The $recalcB()$ and $recalc\sT()$ functions}
\label{s:appendix:alg_recalcs}
%
We present the exact formulas that give the first three moments of $s.B[k]$, i.e., the function $recalcB()$, and $\sT_d$, i.e., the function $recalc\sT()$, where $d= object[k]$. We find these formulas using the equations in sections~\ref{s:blocking}, and respectively,~\ref{s:throughput}.

Let $F_{s,d}$ $=$ $A$ $+$ $R(s,d)$ $\cdot$ $R(d,d)$ $\cdot$ $f_{s,d}$, be the sum of the acquisition time and the job completion period times the related probabilities ($F_{s,d}$ can be bound by a constant and thus can be treated as such).
%
%
The first three moments of the blocking time are
$E(s.B[M]^{m})$ $=$ $(F_{s,M})^m$, for $m$ $=$ $1,2,3$, and $E(s.B[k])$ $=$ $F_{s,d}$ $+$ $\Sigma_{k' = k+1}^M$ $(\Pr[(s$, $\bullet$, $d$, $object[k'])]$ $\cdot$ $E(d.B[k'])$,
%
%
$E(s.B[k]^{2})$ $=$ $(F_{s,object[d]})^2$ $+$ $\Sigma_{k' = k+1}^M$ $\Pr[(s$, $\bullet$, $d$, $object[k'])]$ $(E(d.D[k']^{2})$ $+$ $2F_{s,d}$ $\cdot$ $E(d.D[k']))$ and
%
%
$E(s.B[k]^{3})$ $=$ $(F_{s,d})^3$ $+$ $\Sigma_{k' = k+1}^M$ $\Pr[(s$, $\bullet$, $d$, $object[k'])]$ $(E(d.D[k']^{3})$ $+$ $3(F_{s,d})^2$ $\cdot$ $E(d.D[k'])$ $+$ $3F_{s,d}$ $\cdot$ $E(d.D[k']^{2}))$, if $d=object[k]$ and $k\in [1,M-1]$.
Moreover, the first three moments of $\sT_d$, where $d=object[k]$, are
$E(\sT_d)$ $=$ $(\Sigma_{s \in arrivals(k)}$ $\omega(s$, $d)$ $\cdot$ $(F_{s,d}$ $+$ $E(\sT_s))$ $/$ $(\Sigma_{s \in arrivals(k)}$ $\omega(s$, $d))$,
$E(\sT_d^{2})$ $=$ $(\Sigma_{s \in arrivals(k)}$ $\omega(s$, $d)$ $\cdot$ $((F_{s,d})^2$ $+$ $E(\sT_s^{2})$ $+$ $2F_{s,d}$ $\cdot$ $E(\sT_s))$ $/$ $(\Sigma_{s \in arrivals(k)}$ $\omega(s$, $d))$ and
$E(\sT_d^{3})$ $=$ $(\Sigma_{s \in arrivals(k)}$ $\omega(s$, $d)$ $\cdot$ $((F_{s,d})^3$ $+$ $E(\sT)_s^{3})$ $+$ $3(F_{s,d})^2 E(\sT_s)$ $+$ $3F_{s,d} E(\sT_s^{2}))$ $/$ $(\Sigma_{s \in arrivals(k)}$ $\omega(s$, $d))$. 
We remind that $arrivals(object[k])$ $=$ $thread$ $\cup$ $\{object[j]$ $|$ $j \in [1,k-1]\}$ and $\omega(s$, $object[k])=$ $R(s$, $object[k])\cdot s.T[k]$ (Section~\ref{ss:object_inter_demand_periods}).

\Subsection{Running time}
\label{s:appendix:alg_compl}
Notice that the running times of Algorithm~\ref{alg:framework} (Baynat-Dallery framework) and Algorithm~\ref{alg:yOSEf} depend on the number of iterations of these algorithms. Lemma~\ref{th:running_time_yOSEf} bounds the procedure running time for one iteration.

\begin{lemma}
\label{th:running_time_yOSEf}
The running time of one iteration of Algorithm~\ref{alg:yOSEf} is in $O(M^2\cdot N^4 + M^3)$.
\end{lemma}

\proof{
We look at the running time of each step of the repeat-until loop to find the algorithm's running time.
Steps (i) and (iii) call the function $updateStates()$ so at most $M$; $M$, and respectively $M-1$ times.
Note that the function $updateStates()$ calls at most $1 + (M-1)$ times the function $BDF()$, because the input parameter $k$, which denotes the object whose state is to be updated, is at most $M$ (line~\ref{alg:yOSEf:procedure_for}).
The first call is done by setting $S_k=thread$ and (at most) $M-1$ calls are done by setting $S_k = \{object[j]\}$, $j\in [1,k-1]$.
The $BDF()$ function has running time in $O(M \cdot N^4 \cdot I_f)$ and $O(M\cdot I_f)$ (Lemma~\ref{lem:framework_complexity}), when $BDF()$ is called for $S_k=thread$, and respectively, for $S_k = \{object[j]\}$, $j\in [1,k-1]$, where $I_f$ denotes the maximum number of framework iterations (Lemma~\ref{lem:framework_complexity}).
It also holds that the function $augmntThreadBlock()$ is called $N$ times in step (ii) and its the running time is practically constant (see Appendix~\ref{s:appendix:jobAssignmentProbability} with respect to the findings of Latouche and Ramaswami~\cite{latouche1993logarithmic}).
Thus, the running time of one iteration of Algorithm~\ref{alg:yOSEf} is in $O(M\cdot (M \cdot N^4 \cdot I_f) + N + M\cdot ((M-1) \cdot M \cdot I_f)) = O(M^2\cdot N^4 + M^3)$. 
}


\Section{Conclusions}
We consider a resource allocation problems that can be modeled as generalized dynamic dining philosophers problems.
We formulate questions that are associated with  equilibrium situations in such systems, where input and output rates match.
We believe that the way we find the equilibrium as well as estimate the delay and throughput in such systems can be the basis for an analysis of further generalizations of the problem studied here, such as the ones that are described in the literature on resource allocation, e.g., non-sequential scheduling, such as parallel resource acquisition ($2$-phase locking) and resource acquisition that is reactive to contention conditions~\cite{herlihy2012art, DBLP:books/mk/Lynch96, DBLP:conf/opodis/HaPT04, ha2004reactive, DBLP:conf/opodis/PapatriantafilouT97}.
%


\begin{algorithm*}[h!]
\caption{The procedure for finding an $\varepsilon$-OSE}

\label{alg:yOSEf}

{\bf Input:} Number of objects, $M$ and threads, $N$; jobs $\{job_i\}_{i\in [1,J]}$ and their arrival rates to threads $\{\lambda_{i,n}\}_{n \in [1,N], i \in [1,J]}$; $R$ request probability matrix;

{\bf Variables:}
The item states are recorded in the arrays $thread[1,N]$ and $object[1,M-1]$, such that $thread[n]$ is of the form $\langle \text{(inter-demand period)}~T[1, M]$, $\text{(delay)}~D[1, M]$, (blocking~period)$~B[1, M] \rangle$ and $object[k]$ is of the form  $\langle \text{(inter-demand period)}~T[k+1, M]$, (delay)$~D[k+1, M]$, (blocking~period)$~B[k+1, M] \rangle$; $\sT_v$, item $v$'s inter-demand period, for every $v=thread[n]$ such that $n\in [1,N]$ and $v=object[k]$ such that $k\in [1,M-1]$; $loopEnd$ (Boolean) this variable is true when the function $augmentThreadBlock()$ decides that no OSE can be found and thus the loop should stop.

{\bf Output:} $(thread[1,N], object[1,M-1])$

\noindent{\bf Macros:}
$objThrdSet() = \{\langle tag, i, \tau \rangle | (tag = \#o \land \sT_{object[i]} = \tau) \lor (tag=\#t \land \sT_{thread[i]} = \tau)\}$\label{alg:yOSEf:line:objectThreadSet}
$converged(prev,curr) = (\nexists\, \langle tag, i, \tau \rangle \in prev, \langle tag, i, \tau' \rangle \in curr : |\tau - \tau'|\geq\varepsilon)$\label{alg:yOSEf:line:converged_macro}
$blocking(n)=A + \Sigma_{k = 1}^M R(thread[n], object[k]) \cdot thread[n].D[k]$\label{alg:yOSEf:line:thread_T}
$allStates = \{thread[n] | n\in [1,N]\}\cup \{object[k] | k\in [1,M-1]\}$

\Begin{
$initializeSystemState(N,M,\{job_i\}_{i\in [1,J]}, \{\lambda_{i,n}\})$\label{alg:yOSEf:line:init}

$loopEnd \gets \textbf{ false}$ 

\Repeat{$converged(prevSet, objThrdSet()) \lor loopEnd = \textbf{\em true}$\label{alg:yOSEf:line:convergence}}{\label{alg:yOSEf:line:repeat}
\noindent {\bf let} $prevSet \gets objThrdSet()$ \label{alg:yOSEf:line:loopStart}
\lFor{$k = M$~{\bf to}~$1$}{$updateStates(\#B, k, thread, object, \{\sT_v\}_{v\in allStates}, R)$}\label{alg:yOSEf:line:blocking}
\lForEach{$n \in [1,N]$}{$(\sT_{thread[n]}, loopEnd) \gets augmntThreadBlock(\Sigma_i\lambda_{i,n}, blocking(n))$}\label{alg:yOSEf:line:thread_updates}
\lFor{$k = 1$~{\bf to}~$M-1$}{$updateStates(\#\sT, k, thread, object, \{\sT_v\}_{v\in allStates}, R)$}\label{alg:yOSEf:line:throughput}
}

\Return{$(thread[1,N], object[1,M-1])$}\label{alg:yOSEf:line:until}
}

\noindent{\bf procedure} $initializeSystemState(N,M,\{job_i\}_{i\in [1,J]}, \{\lambda_{i,n}\})$ \label{alg:yOSEf:line:init_start}
\Begin{
\For{$k= M$ {\em to} $1$\label{alg:yOSEf:line:init_s_start}}{\lForEach{item $s$ such that $(s,object[k])$ is an edge in $\dgraph$}{$s \gets initRecord(s, k, thread[1,N], object[1,M-1])$}\label{alg:yOSEf:line:init_s_end}}
\lFor{$n=1$ {\em to} $N$}{$\sT_{thread[n]} \gets augmentThreadBlock(\Sigma_i \lambda_{i,n}, blocking(n))$\label{alg:yOSEf:line:init_T_thread}}
\lFor{$k=1$ {\em to} $M-1$}{$\sT_{object[k]} \gets \Sigma_{\ell = k+1}^M R(object[k], object[\ell]) \cdot object[k].T[\ell]$\label{alg:yOSEf:line:init_T_object}}
}

\noindent{\bf procedure} $updateStates(tag, k, thrSet, objSet, (\sT_{v})_{v\in allStates}, R)$ \label{alg:yOSEf:line:proc_start}
\Begin{
$\cs(thrSet, k) \gets defContentionSubsystem(thread[1,N], object[1,M-1], k, (\sT_{v})_{v\in allStates}, R)$\label{alg:yOSEf:upd_st_def_thr}\;

$(thrSet[n].T[k], thrSet[n].D[k])_{n\in  [1,N]} \gets BDF(\cs(thrSet, k))$\label{alg:yOSEf:line:estCont_thr}\;

\For{$j = 1$~{\bf to}~$k-1$\label{alg:yOSEf:procedure_for}}{

$\cs(\{objSet[j]\}, k) \gets defContentionSubsystem(\{objSet[j]\}, object[1,M-1], k, (\sT_{v})_{v\in allStates}, R)$\label{alg:yOSEf:line:def_obj_cs}

$(objSet[j].T[k], objSet[j][n].D[k]) \gets BDF(\cs(thrSet, k))$\label{alg:yOSEf:line:estCont_obj}

}
\lIf{$tag=\#B$}{$recalcB(k)$, {\bf else if} $tag=\#\sT$ {\bf then} $recalc\sT(k)$}\label{alg:yOSEf:line:proc_end}
}

\end{algorithm*}

\clearpage
\bibliographystyle{abbrv}
\bibliography{localbib}




\end{document}